\documentclass[aps,a4paper,amsmath,amssymb,twocolumn,
floatfix,superscriptaddress,showkeys]{revtex4-1}
\usepackage{bm,amsmath,amssymb,amsfonts}
\usepackage[dvips]{graphicx}
\usepackage[utf8]{inputenc}
\usepackage{ragged2e}
\usepackage[normalem]{ulem}
\usepackage{color}
\usepackage{xcolor}
\usepackage{hyperref}

\hypersetup{
    pdfnewwindow=true,     
    colorlinks=true,       
    linkcolor=magenta,     
    citecolor=blue,        
    filecolor=blue,        
    urlcolor=blue          
}
\begin{document}
\title{Compact localized states and magnetic flux-driven topological 
phase transition\\ in a diamond-dodecagon lattice geometry}
\author{Joydeep Majhi}
\affiliation{AbCMS Lab, Department of Metallurgical Engineering and Materials Science, 
Indian Institute of Technology Bombay, Mumbai, Maharashtra 400076, India} 
\author{Biplab Pal}
\thanks{Corresponding author}
\email[E-mail: ]{biplab@nagalanduniversity.ac.in}
\affiliation{Department of Physics, School of Sciences, 
Nagaland University, Lumami 798627, Nagaland, India}
\date{\today}
\begin{abstract}
We propose and investigate a novel two-dimensional (2D) tight-binding model defined on 
a diamond-dodecagon lattice geometry that hosts multiple flat bands (FBs) and supports 
topological phase transitions driven by a magnetic flux. This lattice exhibits three 
completely flat, non-dispersive bands in the band structure in the absence of magnetic 
flux due to destructive interference in the electron hoppings, leading to the emergence 
of compact localized states (CLS). These CLS are analytically constructed and exhibit 
real-space confinement of the electrons, arising solely due to the lattice's geometrical 
frustration. It has been shown that these FBs are very robust against the introduction 
of weak random onsite disorder in the system. By tuning the uniform magnetic flux 
threaded through the diamond plaquettes, we demonstrate a tunable evolution of the band 
structure and show that certain bands develop nontrivial topological features with nonzero 
integer values of the Chern number. Additionally, we have computed the multi-terminal 
transport properties for this 2D lattice system, which display the flux-tunable resonances 
and transmission suppression linked to the FBs, establishing a clear interplay between 
the localization, topology, and transport. Our findings put forward the diamond-dodecagon 
lattice as a robust and tunable platform for studying the flat-band physics and magnetic 
flux-controlled topological phenomena, offering promising experimental feasibility in 
photonic lattices and ultracold atomic systems. 
\end{abstract}
\maketitle
%

\section{Introduction}
\label{sec:intro}
The exploration of flat-band (FB) systems has gained considerable attention in recent years due to their rich physical 
properties and potential applications in the emergent quantum materials~\cite{FB-Review-Flach-2018,FB-Review-Leykam-2020,
FB-Review-Rodrigo-2021}. Characterized by dispersionless energy eigenstates with zero group velocity and effectively 
infinite mass, flat bands (FBs) arise from the intricate interplay of the lattice geometry and destructive interference in 
hopping pathways. This leads to a macroscopic degeneracy of single-particle states, effectively quenching the kinetic 
energy and allowing interaction effects to dominate. FB systems not only host a wide array of exotic many-body phases, 
such as high-temperature superconductivity~\cite{FB-superconductivity-1,FB-superconductivity-2,FB-superconductivity-3,
FB-superconductivity-4,FB-superconductivity-5}, flat-band ferromagnetism~\cite{FB-ferromagnetism-1,FB-ferromagnetism-2,
FB-ferromagnetism-3,FB-ferromagnetism-4}, and fractional Chern insulators~\cite{FB-Chern-insulator-1,FB-Chern-insulator-2,
FB-Chern-insulator-3}, but also draw fundamental significance from their deep analogy to Landau levels in two-dimensional 
electron gases, which underpin the celebrated integer and fractional quantum Hall effects~\cite{IQHE-Klitzing,IQHE-Laughlin,
FQHE-Laughlin}. From both theoretical and experimental perspectives, FB systems play a pivotal role in understanding 
the synergy between geometry, topology, and electron correlations in engineered lattice systems.

The microscopic origin of FBs in tight-binding lattice models is rooted in destructive quantum interference, which 
suppresses the particle hopping between certain lattice sites and leads to the formation of compact localized states 
(CLS)~\cite{Ajith-prb-2017,Singular-FB-Rhim-2021,Pal-prb-2018,Amrita-prb-2026,Pal-jpcm-2025}. These eigenstates are characterized by 
wavefunctions confined to a finite set of lattice sites, with amplitudes vanishing identically outside this region. 
The existence and structure of CLS are closely tied to the lattice's topology and symmetry, making them a defining 
feature of geometrical frustration~\cite{Sutherland-prb-1986}. Such FB localization mechanisms have been demonstrated 
across a range of geometries, including the Lieb~\cite{FB-in-Lieb,FB-in-Lieb-Kagome-Dice}, 
Kagome~\cite{FB-in-Lieb-Kagome-Dice,FB-in-Kagome}, dice~\cite{FB-in-Lieb-Kagome-Dice,FB-in-Dice}, 
and diamond chain lattices~\cite{FB-in-Diamond-chain-Sebabrata,FB-in-Diamond-chain-Dias,Amrita-jpcm-2021}, as well as 
in various artificially engineered lattices~\cite{Bhatta-Pal-prb-2019,Pal-Saha-prb-2018,Chen-Suepr-Honeycomb-AOM-2020,
Chen-Super-Kagome-OL-2023,Denz-Extended-Lieb-FB,Hahafi-fractal-FB,Chen-fractal-FB-1,Chen-fractal-FB-2}. 
Many of these systems support one or more FBs due to their unique connectivity patterns, and recent experimental 
advances have enabled the physical realization of such FBs in photonic crystals~\cite{Rodrigo-prl-2015,
Sebabrata-prl-2015,Sebabrata-prl-2018,Rodrigo-prl-2022}, ultracold atom systems~\cite{Goldman-pra-2011,Wang-prl-2021,
Longhi-prl-2022}, and moiré heterostructures~\cite{Balents-Moire-FB-1,LeRoy-Moire-FB-2,Li-Moire-FB-3}, 
providing highly tunable platforms for studying FB physics in controlled environments. Very recently, FB lattice 
settings have also been proposed for the realization of robust quantum storage devices~\cite{Rudolf-prl-2026}.

A compelling frontier in the flat-band research is the engineering of band topology through external perturbations. 
Of particular interest is the introduction of magnetic flux through lattice plaquettes~\cite{Pal-prb-2018,Amrita-prb-2026,Pal-jpcm-2025}, 
which modifies the hopping phases via the Peierls substitution and allows the generation of nontrivial topological 
features. These include topological flat bands (TFBs) that possess finite Berry curvature and are associated 
with quantized Chern numbers, akin to Landau levels in a continuum, but without the requirement of a magnetic 
field~\cite{Wen-prl-2011,Das-Sarma-prl-2011,Neupert-prl-2011}. While perfect flatness, finite-range hopping, 
and nontrivial topology cannot coexist in a single band due to theoretical constraints~\cite{Exact-FB-n-topology-n-hopping}, 
TFBs remain promising candidates for hosting fractional Chern insulators -- lattice analogs of fractional quantum 
Hall states, potentially stable at higher temperatures. 
\begin{figure}[ht]
\centering
\includegraphics[clip, width=0.9\columnwidth]{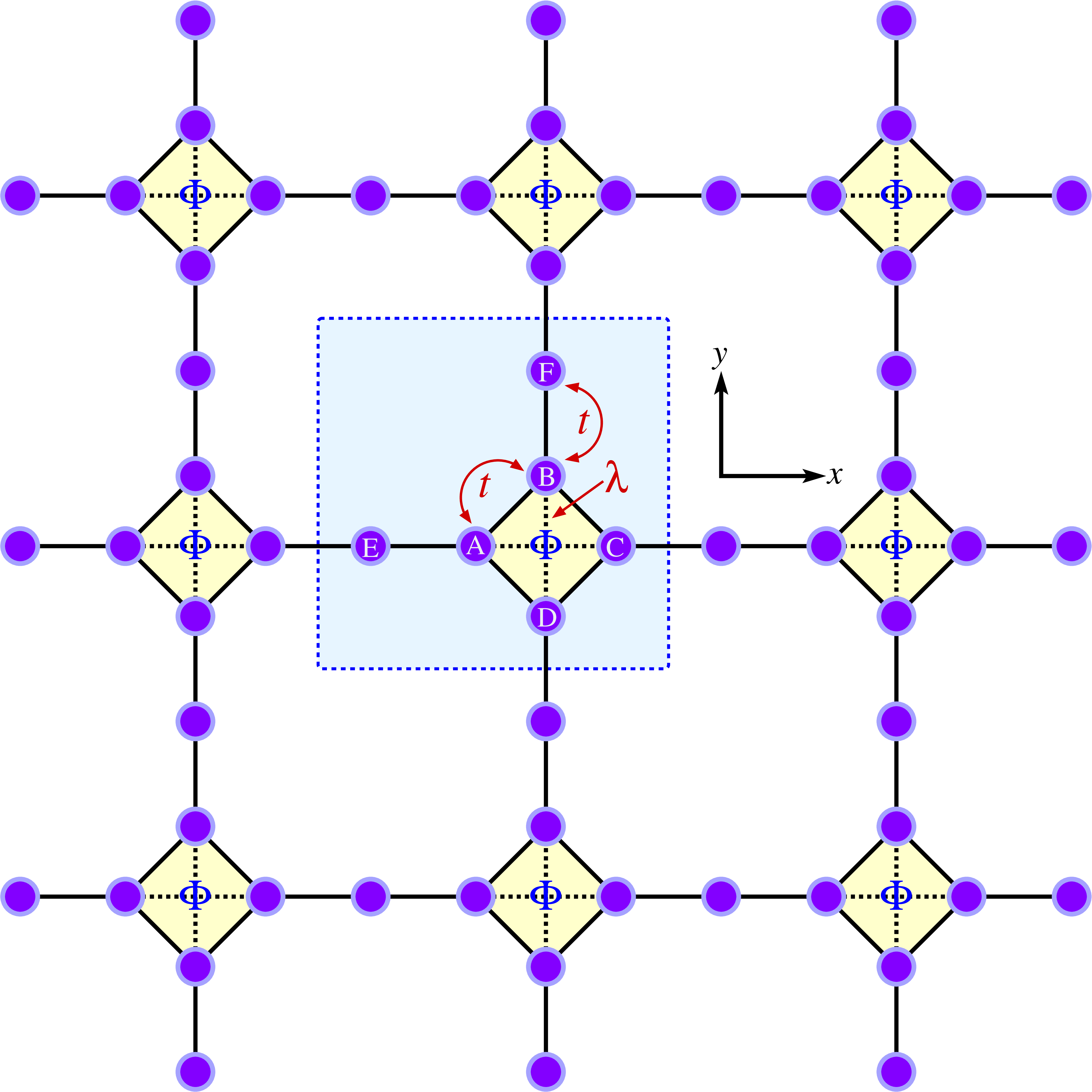}
\caption{Schematic diagram of a two-dimensional (2D) diamond-dodecagon lattice 
model. The unit cell of the lattice containing six atomic sites is highlighted 
with light blue color. Each diamond plaquette is pierced by a uniform magnetic 
flux $\Phi$. $t$ and $\lambda$ represent the hopping parameters 
in the system (see main text).} 
\label{fig:lattice-model}
\end{figure}

In this work, we propose and investigate a novel two-dimensional lattice model, termed as ``diamond-dodecagon lattice", 
which exhibits a rich flat-band structure and nontrivial topological phases. The key findings in this study lies in the 
systematic control of both flat band formation and topological properties through the interplay between an externally 
tunable magnetic flux and the intrinsic lattice hopping parameters. The unit cell of this lattice consists of six 
atomic sites arranged in a 2D geometry composed of interconnected diamond and dodecagon plaquettes. A uniform magnetic 
flux $\Phi$ is introduced through each diamond plaquette (see Fig.~\ref{fig:lattice-model}), which breaks the time-reversal 
symmetry, leading to complex hopping amplitudes governed by the Aharonov-Bohm phase~\cite{A-B-Phase}. The tight-binding 
Hamiltonian includes both nearest-neighbor hopping along the edges of the plaquettes and diagonal hopping inside the 
diamond plaquettes (see Fig.~\ref{fig:lattice-model}), thereby incorporating the geometrical frustration and interference 
effects in a systematic way. We remark that, this 2D lattice model exhibits multiple FBs which is not necessarily true 
for a 1D version of this model or in a conventional Lieb or Kagome lattice. We also note that, this lattice model shows 
more richer spectrum in terms of the FBs and topological phase transitions as compared to its sister lattice, the 
diamond-octagon lattice model~\cite{Pal-prb-2018}. Another interesting feature observed for this lattice model is that, one 
can systematically make any of the six bands in its band structure perfectly flat at will by tuning the value of the external 
magnetic flux $\Phi$. This feature is apparently missing in a diamond-octagon lattice model. 

Our analysis reveals the presence of three completely flat bands for this lattice model in the absence of the magnetic flux, 
associated with CLS that exhibit nontrivial spatial amplitude distributions. These CLS can be analytically constructed and 
are characterized by vanishing wavefunction amplitudes on specific sub-lattices, a hallmark of the destructive interference. 
We demonstrate that the application of magnetic flux leads to a gradual dispersion of the FBs, highlighting the tunability 
of the spectrum via an external control. We further explore the topological character of the bands by computing their Berry 
curvatures and Chern numbers as a function of the magnetic flux. For selected flux values ($\Phi=\Phi_{0}/4$ and $\Phi_{0}/2$), 
we find the emergence of nonzero integer values of the Chern numbers in specific bands, confirming the topological nature of 
these bands. 

To probe the transport properties of the finite system, we construct a scattering geometry by attaching semi-infinite leads 
along the $\pm x$ directions of a $20 \times 20$ diamond-dodecagon lattice. By calculating the two-terminal transmission 
probability as a function of the injection energy and magnetic flux, we observe distinct signatures of FB transport, including 
sharp resonant features and flux-dependent transmission suppression. These results demonstrate the potential of this model 
to act as a flux-tunable quantum transport device. Our findings establish the diamond-dodecagon lattice as a promising 
candidate for realizing and studying FB physics along with controllable topological and transport properties. The model's 
simplicity, combined with its capacity for magnetic tunability and the presence of robust compact localized states, makes 
it an attractive framework for the experimental implementation in photonic as well as in ultracold atomic systems. 
Moreover, the interplay between flatness, topology, and transport uncovered for this lattice model offers a deeper insight 
into the design principles necessary for engineering topological FB materials.

The remainder of this paper is organized as follows: Sec.~\ref{sec:model} introduces the tight-binding Hamiltonian of the 
model and presents a detailed analysis of the band structure as a function of the magnetic flux and the hopping parameters. 
Sec.~\ref{sec:CLS} demonstrates the explicit construction of the CLS corresponding to the FBs and analyzes their spatial 
structure along with the computation of the density of states. Sec.~\ref{sec:topology} examines the topological properties 
of the model through systematic calculations of the Berry curvatures and Chern numbers. Sec.~\ref{sec:transport} 
investigates the transport properties of the finite system with attached leads, revealing the signatures of the FB physics 
in the transmission spectra. Finally, in Sec.~\ref{sec:conclu}, we conclude with a summary of our key findings and an outlook 
on the future directions, including the role of interactions and potential applications in quantum devices.

\section{The lattice model and the spectrum}
\label{sec:model}
\subsection{Geometry and Real-Space Hamiltonian}
We investigate a 2D tight-binding model defined on a periodic lattice composed of interconnected diamond and dodecagon 
plaquettes, hence referred to as the \textit{diamond-dodecagon} lattice. The unit cell consists of six atomic sites, 
labelled as $A, B, C, D, E,\: \text{and}\: F$, as illustrated in Fig.~\ref{fig:lattice-model}. Each diamond plaquette 
is threaded by a uniform magnetic flux $\Phi$, which induces a Peierls phases in the hopping amplitudes along the arms 
of the diamond plaquettes. This magnetic flux serves as a crucial tuning parameter that allows systematic control over 
both the band structure and topological properties of the system. The dodecagon plaquettes provide additional connectivity 
that enhances the geometric frustration, while maintaining the overall lattice periodicity. The lattice extends periodically 
along the both $x$- and $y$-directions. 

The system includes two types of hopping parameters:
the nearest-neighbor hopping, denoted by $t$, which occurs along the arms of the diamond and dodecagon plaquettes (shown 
by solid black lines in Fig.~\ref{fig:lattice-model}), and the next-nearest-neighbor hopping, denoted by $\lambda$, which 
connects the two opposite sites across each diamond plaquette (shown by dotted lines in Fig.~\ref{fig:lattice-model}).
The electronic properties of the diamond-dodecagon lattice are described by a tight-binding Hamiltonian written in the 
Wannier basis as, 
\begin{widetext}
\begin{equation}
\bm{\hat{H}} = \sum_{m,n} \bigg[ 
\underbrace{\sum_{i} \varepsilon_i c_{m,n,i}^{\dagger} c_{m,n,i}}
_{\textbf{\textcolor{blue}{Onsite potential}}} +
\underbrace{\sum_{i,j} \Big(\mathcal{T}_{ij} c_{m,n,i}^{\dagger} c_{m,n,j} + \text{h.c.}\Big)}
_{\textbf{\textcolor{blue}{Intra-cell hopping}}} +
\underbrace{\sum_{ i,j} \Big( \Lambda_{ij} \big[ c_{m,n,i}^{\dagger} c_{m+1,n,j} + c_{m,n,i}^{\dagger} c_{m,n+1,j}\big]  
+ \text{h.c.} \Big)}_{\textbf{\textcolor{blue}{Inter-cell hopping}}}
\bigg],
\label{eq:real-space-hamiltonian}
\end{equation}
\end{widetext}
where $(m,n)$ denotes the unit cell indices, and $(i,j) \in \{A,B,C,D,E,F\}$ represent the atomic site indices within each 
unit cell. The operators $c_{m,n,i}^{\dagger}$ and $c_{m,n,i}$ are the creation and annihilation operators, respectively, 
for a spinless and non-interacting electron at site $i$ in the $(m,n)$-th unit cell. $i,j$ indicates the nearest-neighbor 
or next-neatest-neighbor pairs depending on the position of the lattice sites as explained below in details. 
$\varepsilon_{i}$ is the onsite potential of an electron at the $i$-th atomic site, and for our model, 
we set the onsite potentials to be identical across all sites: $\varepsilon_i = 0 \ \text{for all } i \in \{A,B,C,D,E,F\}$. 
We remark that, later in Sec.~\ref{sec:CLS}, we also explore the effect of an random onsite potential disorder on the FBs in 
this model. 

As prescribed in Eq.~\eqref{eq:real-space-hamiltonian}, we have two different types of hopping parameters for our lattice 
model, namely, the intra-cell hopping (within a unit cell) and the inter-cell hopping (one unit cell to another unit cell). 
The intra-cell hopping amplitude takes the values:
\begin{equation*}
\mathcal{T}_{ij} = \begin{cases}
t e^{\pm i\theta} & \text{for the diamond edges ($+$ for clockwise } \\
                  & \text{and $-$ for counterclockwise hopping),} \\
\lambda & \text{for diagonal hopping within a diamond } \\
        & \text{(when $i$ and $j$ are next-nearest neighbors),} \\
t & \text{for hopping along the dodecagon edges.}
\end{cases}
\label{eq:intra_hopping}
\end{equation*}
where $\theta$ is the Aharonov-Bohm phase factor given by, $\theta = {2\pi\Phi}/{4\Phi_0}$; 
$\Phi_{0} = h/e$ being the fundamental flux quantum. \\
The inter-cell hopping amplitude takes the values:
\begin{equation*}
\Lambda_{ij} = \begin{cases}
t & \text{for nearest-neighbor connections between } \\ 
  & \text{the adjacent unit cells,} \\
0 & \text{otherwise.}
\end{cases}
\label{eq:inter_hopping}
\end{equation*}
Throughout our calculation, we set $t=1$, and the energy of the electrons $E$ is measured in units of $t$. 
\begin{figure*}[ht]
\centering
\includegraphics[clip, width=0.325\textwidth]{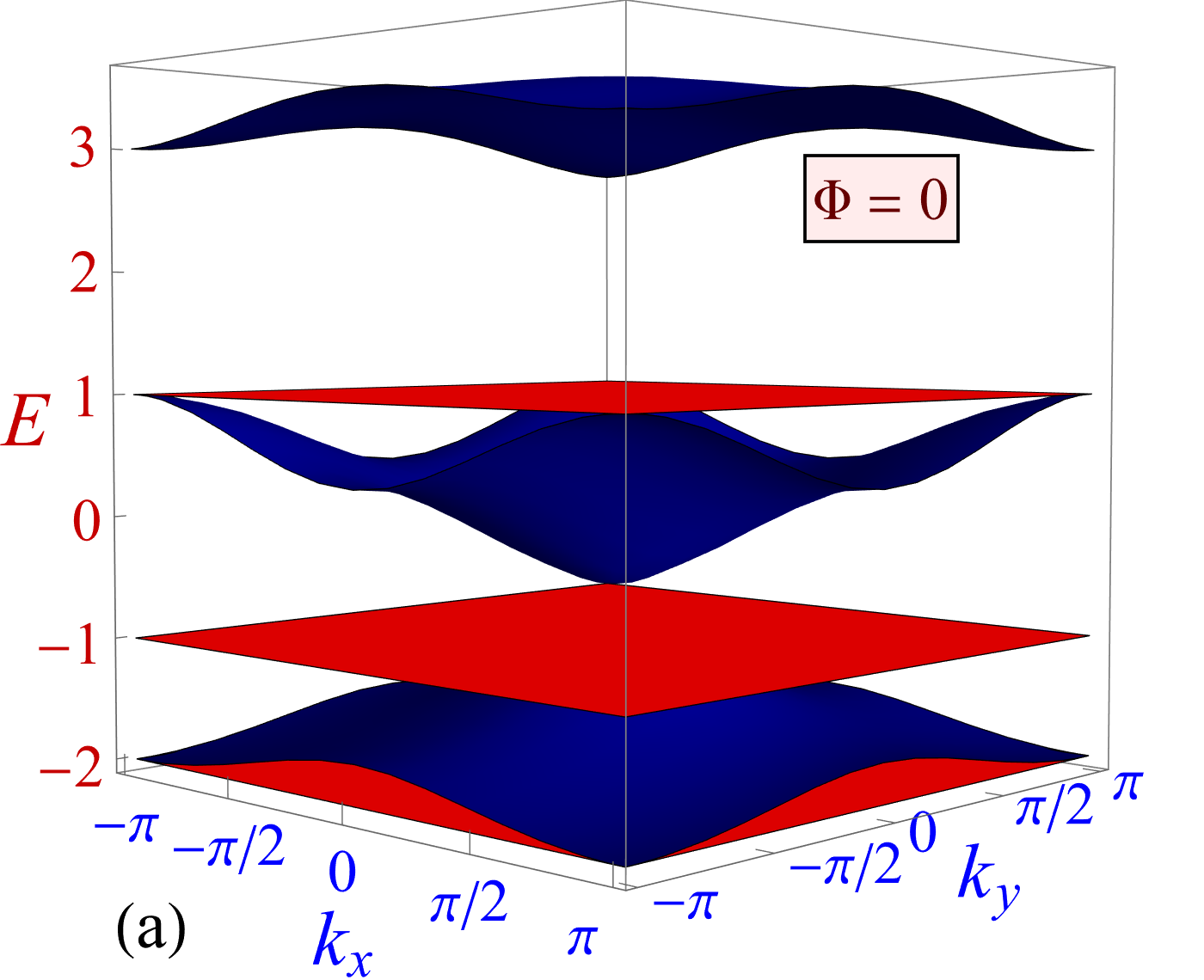}
\includegraphics[clip, width=0.325\textwidth]{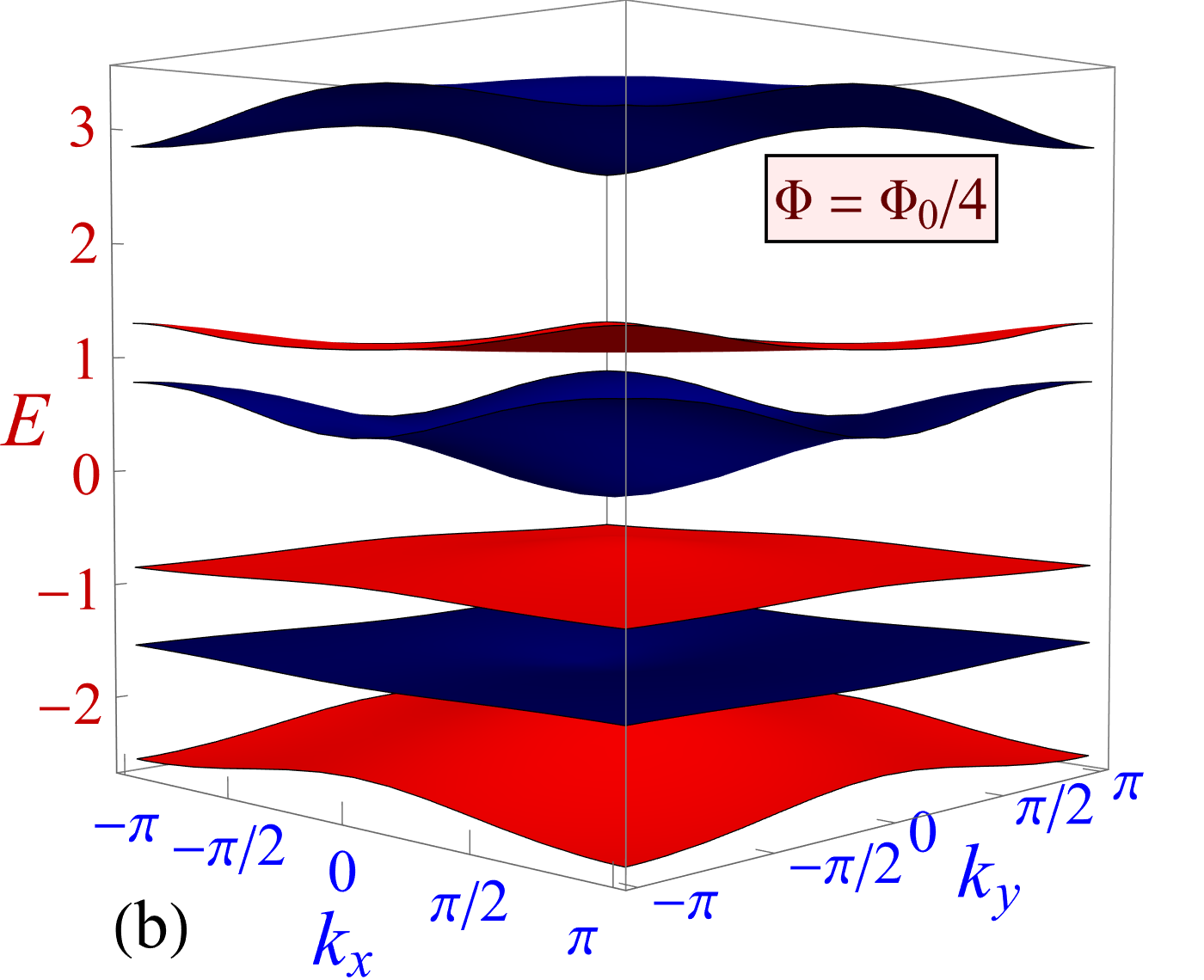}
\includegraphics[clip, width=0.325\textwidth]{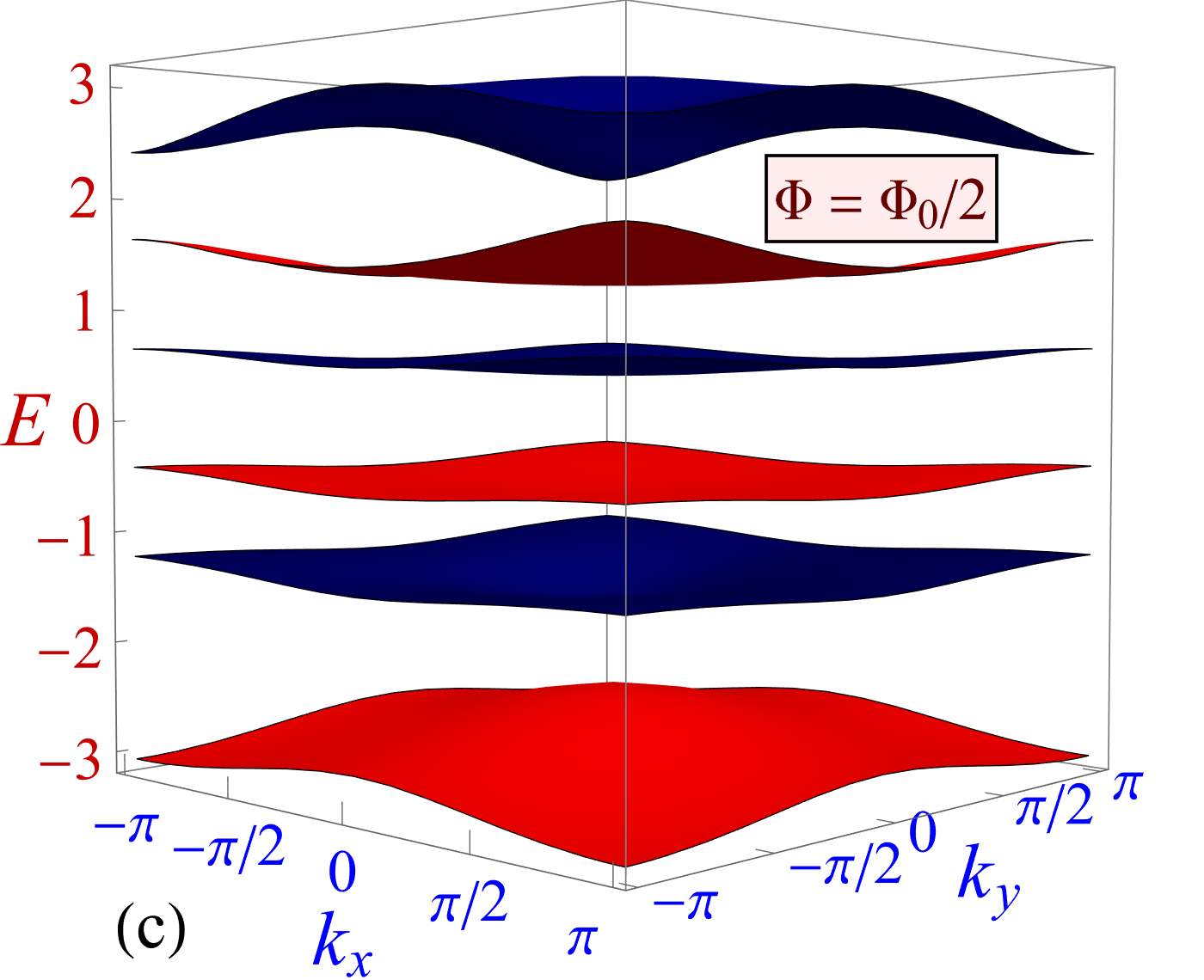}
\caption{Electronic band structure for a 2D diamond-dodecagon lattice for three 
different values of the magnetic flux, \textit{viz.}, (a) $\Phi=0$, 
(b) $\Phi=\Phi_{0}/4$, and (c) $\Phi=\Phi_{0}/2$. For $\Phi=0$, we have three 
completely flat, non-dispersive bands appearing in the system marked by the red 
color, while the remaining three dispersive bands are shown in blue color. 
For $\Phi \ne 0$, gap opens up between all the bands in the system, and 
the FBs are no longer completely flat; they acuire quasi-flatness.
We have set $\varepsilon_{i}=0$ for all sites and $t = \lambda = 1$.}
\label{fig:band-structure}
\end{figure*}

\subsection{Bloch Hamiltonian}
To analyze the band structure of this lattice model, we transform the real-space Hamiltonian in 
Eq.~\eqref{eq:real-space-hamiltonian} to momentum space using the discrete Fourier transform:
\begin{equation}
c_{m,n,j} = \frac{1}{\sqrt{N}} \sum_{\bm{k}} c_{\bm{k},j}\: e^{i\bm{k} \cdot \bm{R}_{m,n}},
\label{eq:Fourier}
\end{equation}
where $\bm{R}_{m,n}$ is the position vector of the $(m,n)$-th unit cell, $N$ is the total number of unit cells, and the sum 
runs over all allowed $\bm{k}$ points in the first Brillouin zone. The resulting momentum-space Hamiltonian takes the form: 
\begin{equation}
\bm{\hat{H}} = \sum_{\bm{k}} \Psi_{\bm{k}}^{\dagger} \mathcal{H}(\bm{k}) \Psi_{\bm{k}},
\label{eq:Bloch-hamiltonian}
\end{equation}
where the Bloch spinor is defined as:
\begin{equation}
\Psi_{\bm{k}}^{\dagger} = \left( c_{\bm{k},A}^{\dagger} \quad c_{\bm{k},B}^{\dagger} 
\quad c_{\bm{k},C}^{\dagger} \quad c_{\bm{k},D}^{\dagger} 
\quad c_{\bm{k},E}^{\dagger} \quad c_{\bm{k},F}^{\dagger} \right).
\label{eq:Bloch-spinor}
\end{equation}
The $6 \times 6$ Bloch Hamiltonian matrix $\mathcal{H}(\bm{k})$ is given by:
\begin{align}
\hspace{-0.15cm}
\mathcal{H}(\bm{k}) = 
\left(\def\arraystretch{1.5} \begin{matrix}
0 & t e^{i\theta} & \lambda & t e^{-i\theta} & t & 0 \\
t e^{-i\theta} & 0 & t e^{i\theta} & \lambda & 0 & t \\
\lambda & t e^{-i\theta} & 0 & t e^{i\theta} & t e^{ik_x} & 0 \\
t e^{i\theta} & \lambda & t e^{-i\theta} & 0 & 0 & t e^{-ik_y} \\
t & 0 & t e^{-ik_x} & 0 & 0 & 0 \\
0 & t & 0 & t e^{ik_y} & 0 & 0 \\
\end{matrix}\right).
\label{eq:Bloch-matrix}
\end{align}

\subsection{Band Structure Analysis}
Using the Bloch Hamiltonian in Eq.~\eqref{eq:Bloch-matrix}, we analyze the evolution of the electronic band structure of 
the diamond-dodecagon lattice both in the absence and in the presence of magnetic flux $\Phi$. Fig.~\ref{fig:band-structure} 
presents the band dispersion $E(k_x, k_y)$ for three representative values of the flux: $\Phi = 0$, $\Phi = \Phi_0/4$, and 
$\Phi = \Phi_0/2$, highlighting the role of quantum interference, localization, and flux-induced symmetry breaking in the 
system. The energy eigenvalues of the system are obtained by solving the eigenvalue equation: 
\begin{equation}
\det\left[\mathcal{H}(\bm{k}) - E_n(\bm{k})\mathbb{I}\right] = 0.
\label{eq:eigenvalue}
\end{equation}
This yields six energy bands $E_n(\bm{k})$ with $n = 1, 2, \ldots, 6$ as functions of the 
crystal momentum $\bm{k} \equiv (k_x, k_y)$ within the first Brillouin zone.

We first discuss the case of $\Phi=0$. In the absence of magnetic flux $\Phi$, the system preserves time-reversal symmetry. 
The resulting band structure exhibits three perfectly flat bands at the energies $E_{FB} = -2$, $-1$, and $+1$ (see 
Fig.~\ref{fig:band-structure}(a)), characterized by zero group velocity ($\nabla_{\bm{k}} E_{n} = 0$) across the first 
Brillouin zone. The other three bands are dispersive, spanning the ranges $E \in [-2, -1]$, $[-1, 1]$, and $[2, 3]$, 
respectively. These FBs arise from the destructive interference in the hopping pathways, leading to compact localized states 
(CLS), whose wavefunctions are confined to finite regions. The eigenstates of the dispersive bands are extended and contribute 
to the bulk transport properties, in contrast to the CLS-based flat bands. 

Upon introducing a magnetic flux $\Phi \ne 0$, the hopping amplitude $t$ acquires a complex Peierls phase 
$t \rightarrow t e^{\pm i\theta}$, breaking the time-reversal symmetry and altering the interference conditions. 
As a result, all the bands are now gapped out from each other, and the perfect flatness of the three FBs does not 
hold anymore, as they become quasi-flat. This partial delocalization indicates the onset of flux-induced kinetic mixing. 
The above observations are true for any nonzero values of the flux; for example, here we have shown two such cases -- 
one is for $\Phi=\Phi_{0}/4$ and the other one is for $\Phi=\Phi_{0}/2$ (see Fig.~\ref{fig:band-structure}(b) and (c)). 
Furthermore, under this condition, topological properties of the system begin to emerge: some of the bands now acquire 
finite integer values of the Chern numbers, reflecting the broken time-reversal symmetry and a redistribution of the 
Berry curvature across the Brillouin zone, which is discussed in detail in Sec.~\ref{sec:topology}. This regime marks 
the transition from the geometrically protected completely FBs to the topologically nontrivial nearly FBs. 
\begin{figure*}[ht]
\centering
\includegraphics[clip, width=0.285\textwidth]{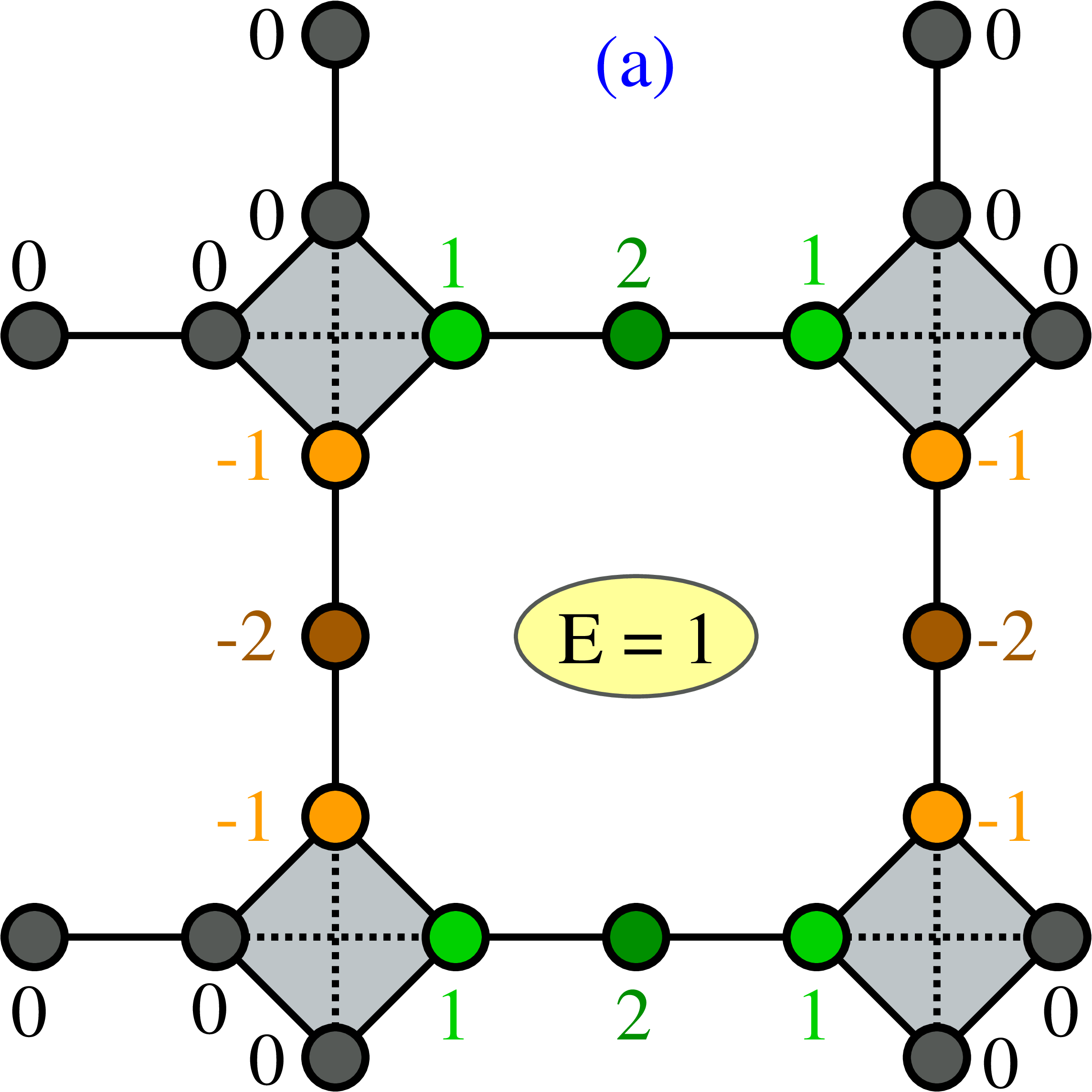}
\qquad
\includegraphics[clip, width=0.285\textwidth]{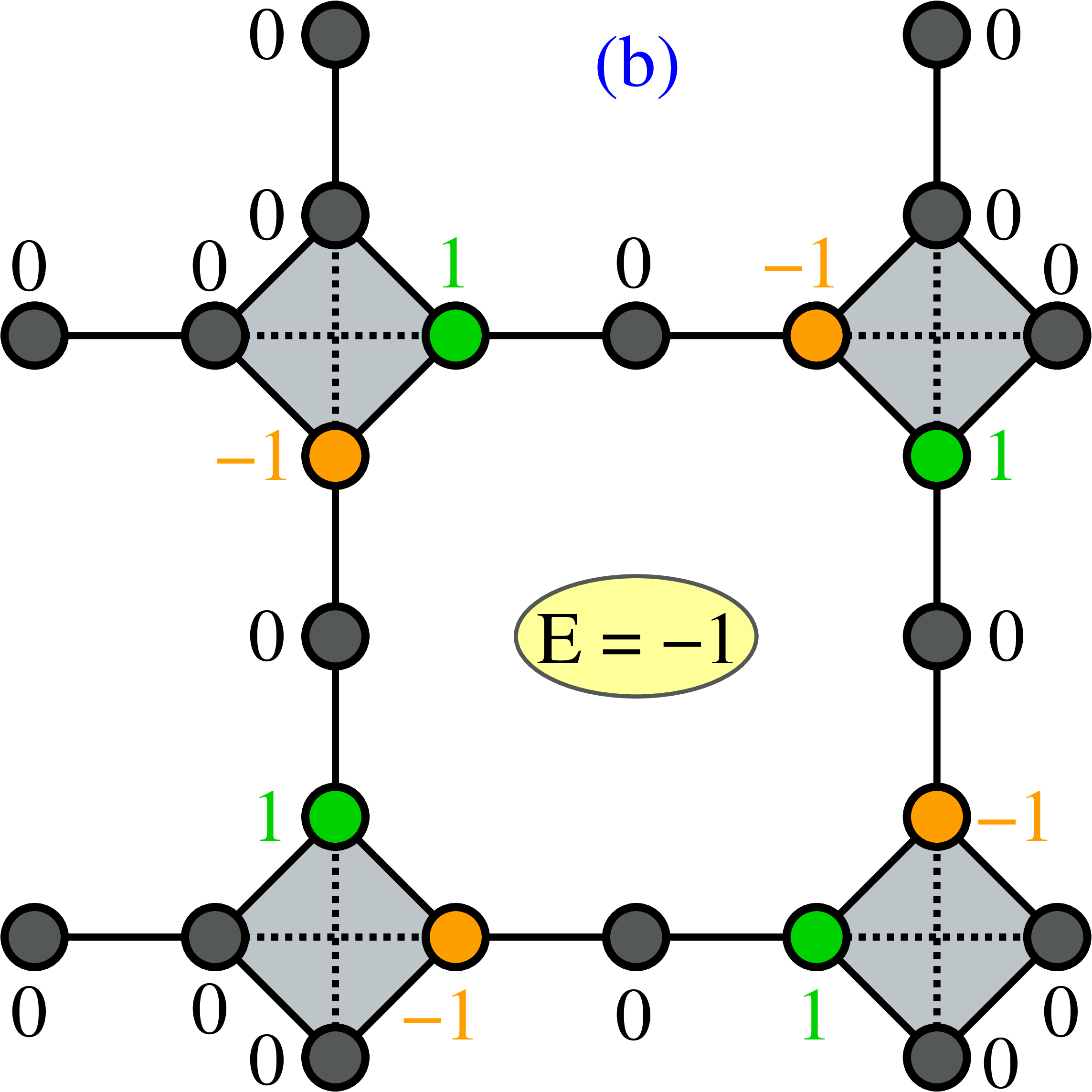}
\qquad
\includegraphics[clip, width=0.285\textwidth]{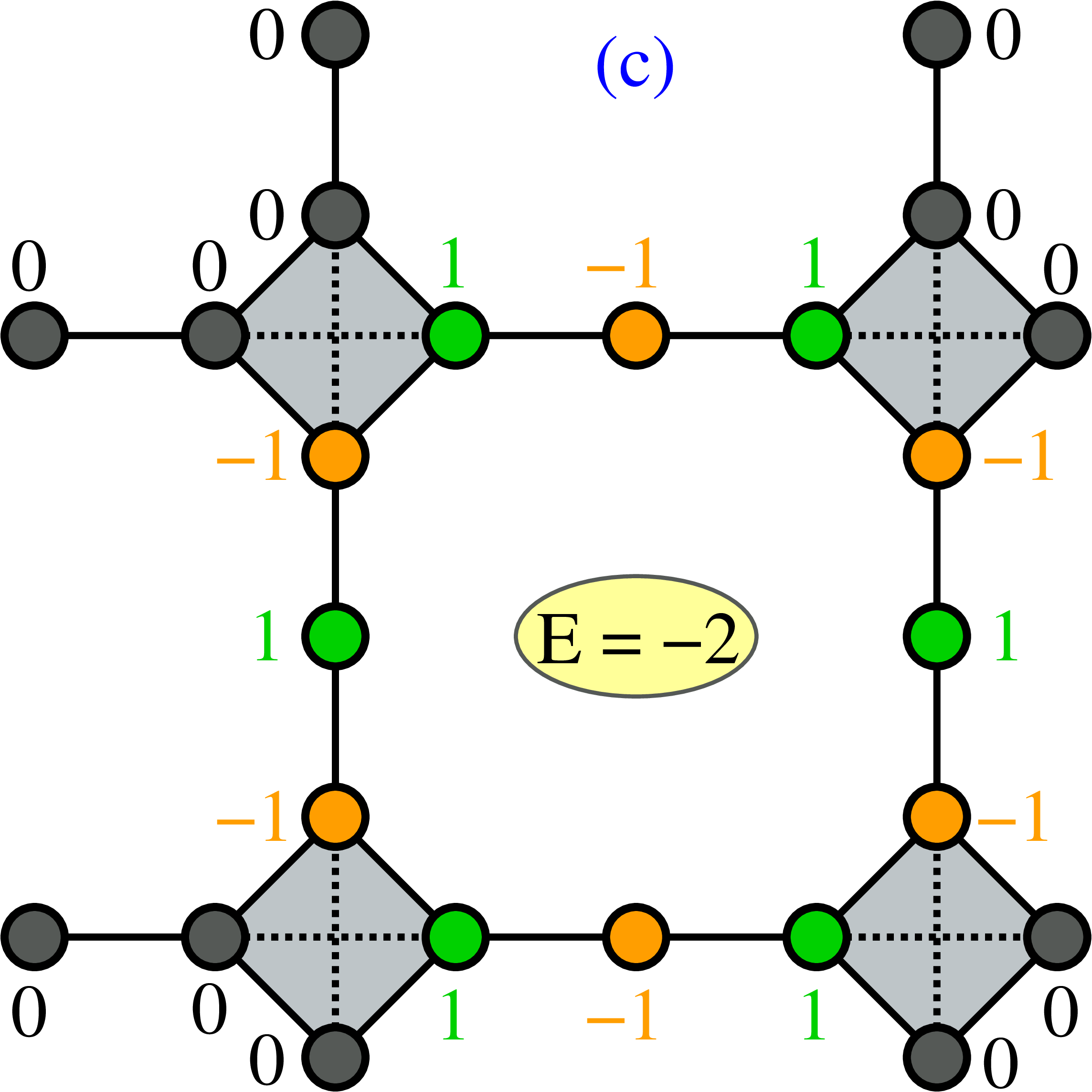}
\caption{Distribution of the compact localized states (CLS) corresponding to the 
flat bands at the energies (a) $E = 1$, (b) $E = -1$, and (c) $E = -2$ in the 
absence of magnetic flux. The black sites denote the nodes with zero wavefunction 
amplitudes, while the dark and light green sites indicate nonzero positive 
wavefunction amplitudes (normalized to $+2$ and $+1$, respectively), and the dark 
and light orange sites indicate nonzero negative wavefunction amplitudes 
(normalized to $-2$ and $-1$, respectively).}
\label{fig:CLS}
\end{figure*}

The flux-tunable band structure of the diamond-dodecagon lattice demonstrates how magnetic flux can be used to engineer 
flatness, break the time-reversal symmetry, and induce the topological phases. Such tunable features open the pathways 
towards realizing the fractional Chern insulators and interaction-driven quantum phases in a diamond-dodecagon lattice 
model in experimentally accessible platforms, such as photonic lattices and cold atom systems. Before ending this section, 
we would like to highlight an interesting feature observed in the band structure of this lattice model as a function of 
the magnetic flux $\Phi$. We have found that, one can make any of the six bands of the system completely flat at will 
for certain specific values of the magnetic flux $\Phi$. This is discussed in the Appendix~\ref{Appendix-A}.

\section{Flat bands and compact localization}
\label{sec:CLS}
In a translationally invariant system, one typically expects to have the Bloch states with extended wavefunctions. 
However, under specific conditions dictated by the lattice geometry and the hopping parameters, destructive quantum 
interference can trap the particles at certain specific values of the energy within a finite region of the 
corresponding lattice. These trapped states, known as the \emph{compact localized states} (CLS), have wavefunctions 
that vanish identically outside a finite support region. The existence of CLS is a hallmark of a flat-band system. 
In this section, we will analyze the structure and formation mechanism of the CLS in the diamond-dodecagon 
lattice model, corresponding to the three exactly FBs observed in the zero-flux band structure (see 
Fig.~\ref{fig:band-structure}(a)). These states originate from the geometry-induced destructive quantum interference 
in the particle hopping pathways and offer deep insight into the microscopic mechanism of FB formation from the 
real-space point of view. 

Fig.~\ref{fig:CLS} presents the real-space wavefunction amplitude distributions of the three CLS corresponding to the FB 
energies $E_{FB} = +1$, $-1$, and $-2$ respectively. Each CLS is strictly confined to a small cluster of sites forming 
a closed loop, which confirms the localized and interference-driven origin of these states. These clusters consist of 
sites with nonzero wavefunction amplitudes, normalized to $\pm 2$ and $\pm 1$, respectively (shown in dark and 
light green and orange colors, respectively in Fig.~\ref{fig:CLS}(a)-(c)), and they are decoupled from another such 
cluster with nodes having zero wavefunction amplitudes (shown in black color in Fig.~\ref{fig:CLS}(a)-(c)). 
We point out that, the CLS amplitude distribution corresponding to the three energies, $E_{FB} = +1$, $-1$, and $-2$ 
have their own distinct patterns. Such variation in their amplitude distributions underscores that 
different FBs may arise from distinct interference patterns and local symmetries. 

The CLS amplitude distributions have been obtained using the following 
difference equation satisfied consistently for all the atomic sites: 
\begin{equation}
\left(E-\varepsilon_{i}\right) \psi_{i} = \sum_{j}\left(\mathcal{T}_{ij} + \Lambda_{ij}\right) \psi_{j},
\label{eq:difference-eqn} 
\end{equation}
with the additional constraint that $|\psi_{\text{CLS}}(\bm r)|^2 = 0$ outside a finite compact region. Each CLS can be 
seen as a local eigenmode supported on a small cluster of lattice sites, with vanishing overlap to the neighboring clusters.  
The formation of CLS has profound physical implications. In momentum space, these localized states contribute equally to 
all $\bm k$-points, resulting in the characteristic flat dispersion. This can be understood from their Fourier transformation 
relationship: a state perfectly localized in the real space must have a uniform distribution in the momentum space. The FB 
states are thus fundamentally different from the dispersive band states. While typical Bloch states carry current and contribute 
to the transport, CLS are immobile by construction. Particles in these states are effectively ``caged" by destructive 
interference, leading to an infinite effective mass and zero group velocity. Unlike Anderson localization, which requires 
disorder, or Wannier-Stark localization, which needs an external electric field, the CLS in our system emerge due to 
the intrinsic structure of the lattice. This makes them robust against small perturbations that preserve the underlying 
symmetries, such as the $x$- and $y$- mirror symmetries, as well as the diagonal mirror symmetries. 
Similar compact localized modes have been extensively studied in other FB lattice models, such as the Lieb,
Kagome, and diamond-octagon lattices. In particular, the role of the interference-induced localization is well-established 
in these systems, making the diamond-dodecagon lattice a compelling addition to the family of CLS-hosting geometries.

To further characterize the spectral features of the diamond-dodecagon lattice and validate the existence of flat bands, 
we compute the average density of states (ADOS) using the retarded Green’s function formalism. The ADOS is a crucial quantity 
that reveals how the energy levels are distributed across the spectrum, offering insight into the presence of the dispersionless 
states and their robustness against disorder. The ADOS $\rho(E)$ is calculated numerically using the standard expression: 
\begin{equation}
\rho(E) = -\frac{1}{\mathcal{N} \pi} \, \operatorname{\Im} \left[ \mathrm{Tr} \, \bm{G}(E) \right],
\label{eq:ADOS}
\end{equation}
where $\bm{G}(E)$ is the Green’s function matrix defined by,
\begin{equation}
\bm{G}(E) = \left[(E + i\eta)\bm{I} - \bm{H} \right]^{-1}.
\end{equation}
Here, $\mathcal{N}$ denotes the total number of lattice sites, $\bm{I}$ is the $\mathcal{N} \times \mathcal{N}$ identity matrix, 
$\bm{H}$ is the full tight-binding Hamiltonian matrix in real space, and $\eta$ is a small broadening parameter (taken to be 
$\sim 10^{-3}$ in our numerical simulations) ensuring convergence and mimicking finite-lifetime effects. The imaginary part of 
the trace of $\bm{G}(E)$ captures the total spectral weight at the energy $E$. 
\begin{figure}[ht]
\centering
\includegraphics[clip, width=0.49\columnwidth]{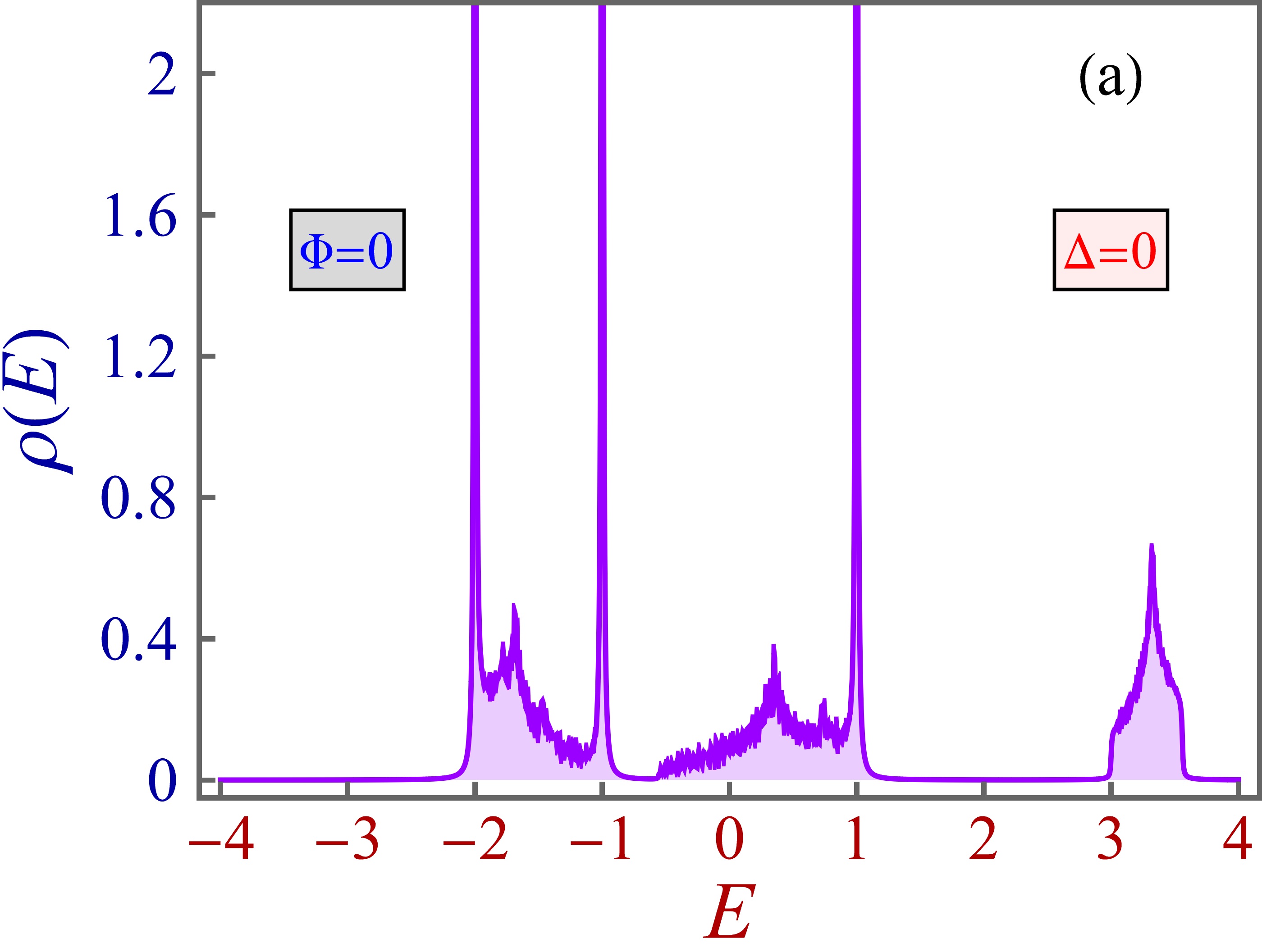}
\includegraphics[clip, width=0.49\columnwidth]{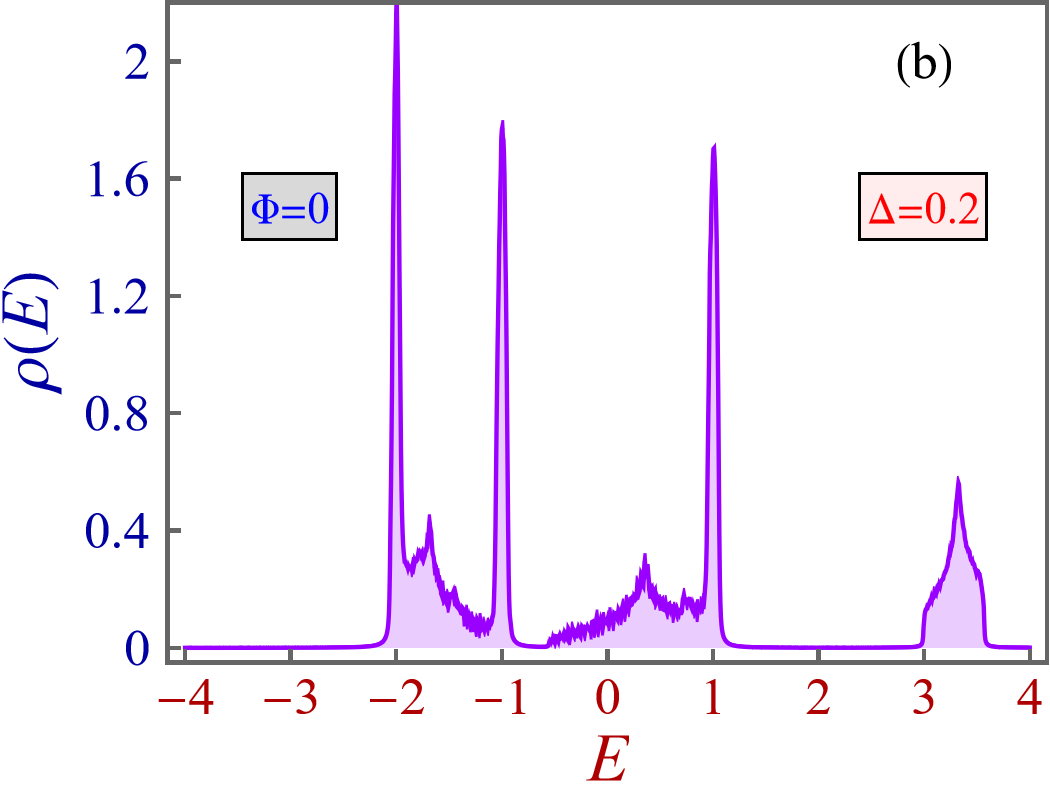} \\
\vskip 0.2cm
\includegraphics[clip, width=0.49\columnwidth]{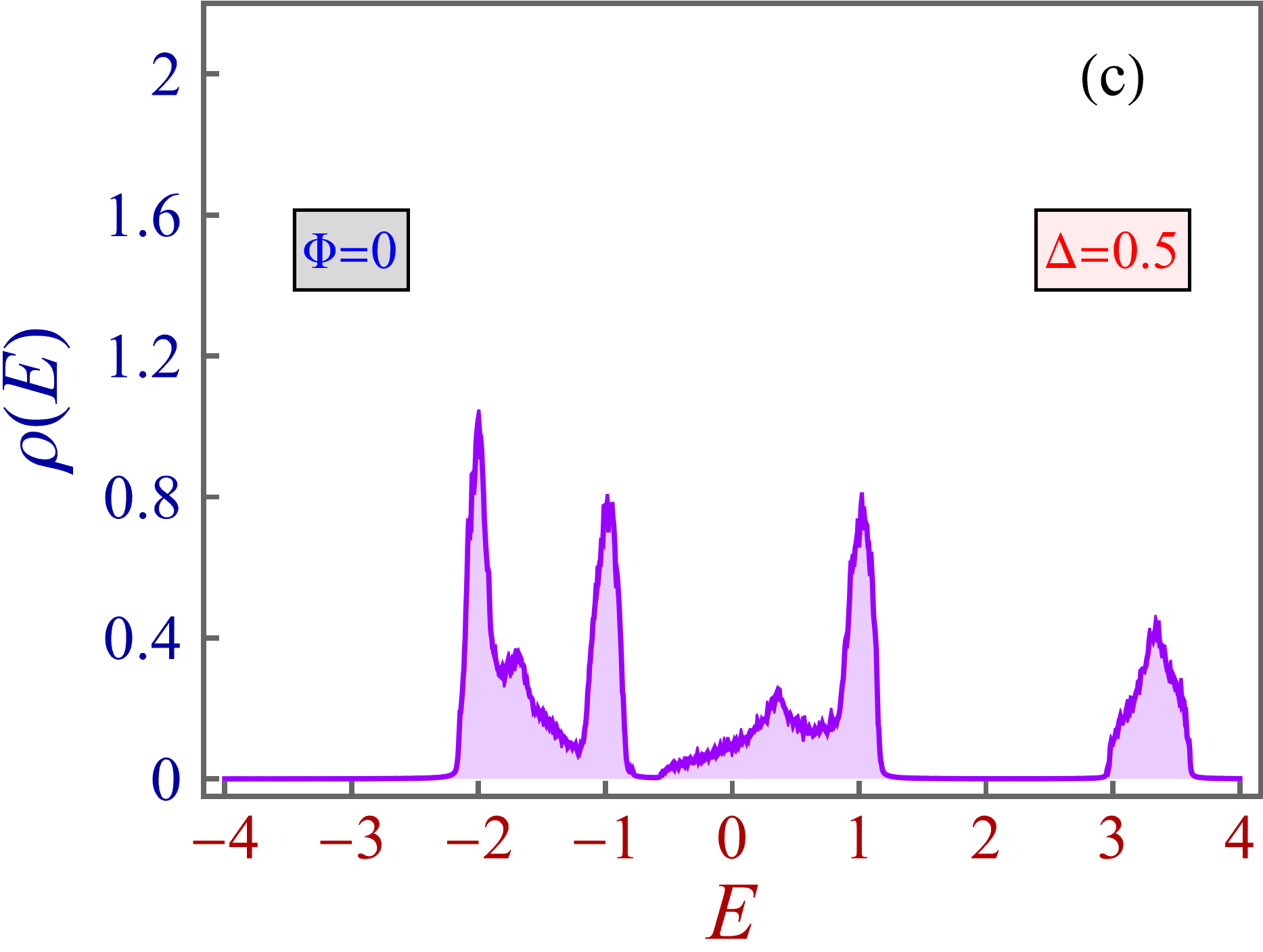}
\includegraphics[clip, width=0.49\columnwidth]{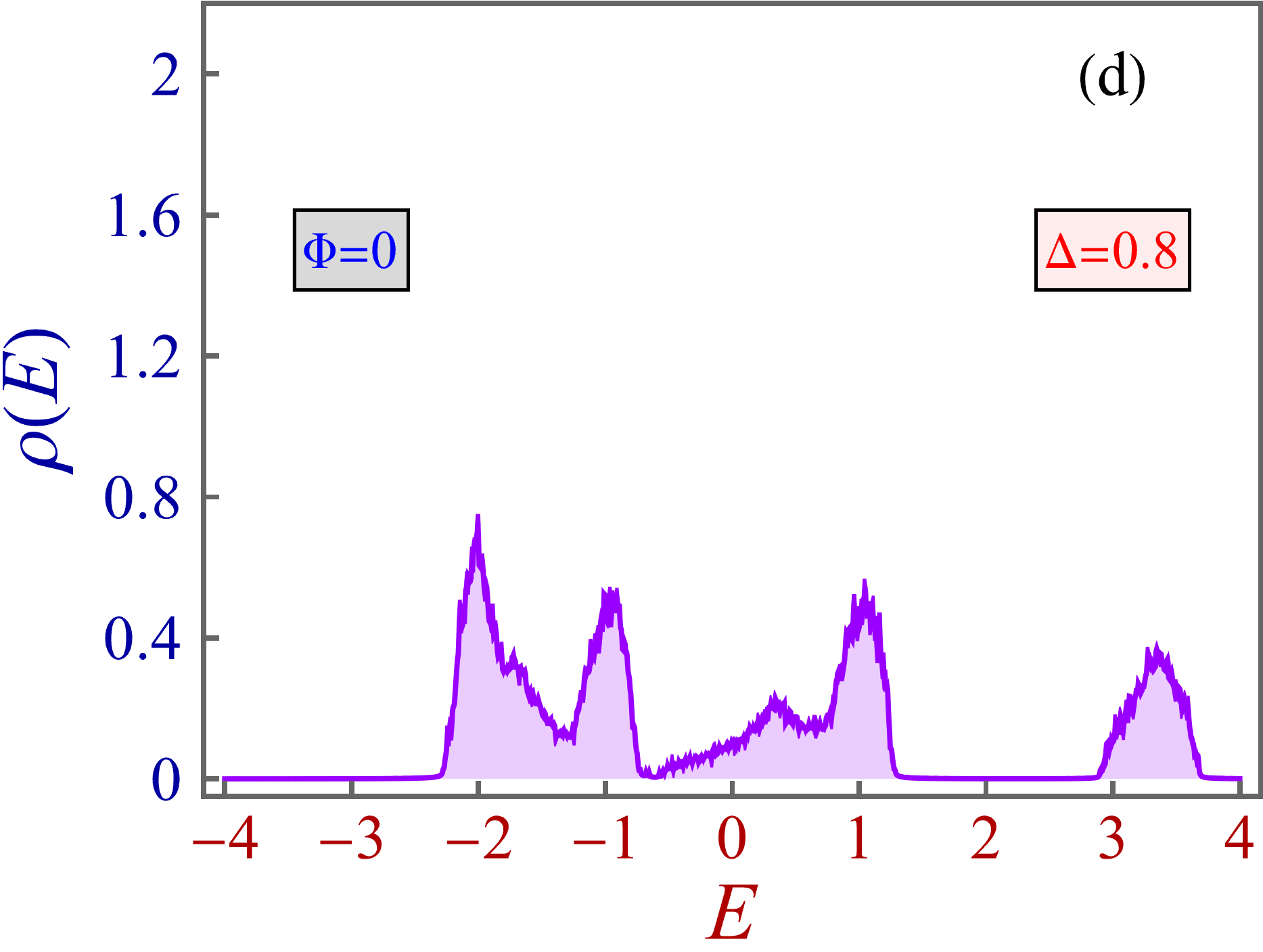}
\caption{The variation of the average density of states (ADOS) $\rho(E)$ of the 
system as a function of the energy ($E$) for $\Phi=0$. (a) is for the perfectly 
ordered system ($\Delta=0$); The sharp peaks in the spectrum indicate the flat 
band states. (b) is for a random disordered distribution of the onsite energies 
with a disorder strength $\Delta=0.2$ (measured in units of $t$). (c) and (d) are 
same as (b) with disorder strength $\Delta=0.5t$ and $0.8t$, respectively. 
The system size considered here is $30 \times 30$ unit cells (\textit{i.e.}, 
$\mathcal{N}=5400$ sites).}
\label{fig:DOS}
\end{figure}

Fig.~\ref{fig:DOS} displays the ADOS computed for the two scenarios: (a) corresponds to the perfectly ordered lattice 
($\Delta = 0$) and (b) shows the disordered case with onsite disorder strength $\Delta = 0.2$, where the onsite energies 
are chosen randomly from a uniform distribution in the interval $[-\Delta/2, \Delta/2]$. 

(a) \textit{Ordered System} ($\Delta = 0$):
In the absence of any disorder in the system, the ADOS reveals three sharp delta-function-like peaks at $E = -2$, $E = -1$, 
and $E = +1$ (see Fig.~\ref{fig:DOS}(a)). These peaks are the direct signatures of the three perfectly flat bands present in 
the band structure (cf. Fig.~\ref{fig:band-structure}(a)). The delta-function-like nature of these peaks arises due to 
the macroscopic degeneracy and zero bandwidth of the FBs, resulting in a diverging density of states. Each flat band 
contributes to a singularity in $\rho(E)$, indicating that a macroscopic number of states are concentrated at a single 
energy level. The appearance of these sharp peaks in the ADOS spectrum corresponding to the FB energies corroborates the 
CLS constructions, in which the nonzero wavefunction amplitudes are localized over small clusters of lattice sites in 
the real-space. Apart from the sharp FB peaks, the ADOS also shows broader continuum structures, which correspond 
to the dispersive bands. These regions reflect the energy intervals over which the extended Bloch states are spread.

(b) \textit{Disordered System} ($\Delta = 0.2$):
Introducing a finite random onsite disorder in the system breaks the perfect translational invariance and perturbs 
the FB states. Here, we have chosen a random onsite disorder of strength $\Delta = 0.2$ (measured in units of $t$). 
As a result, the delta-function-like peaks in the ADOS spectrum start to broaden into finite-width peaks 
(see Fig.~\ref{fig:DOS}(b)). However, the spectral remnants of the FBs are still visible. Physically, disorder introduces 
hybridization between the CLS and the extended states, lifting the perfect degeneracy and leading to a finite dispersion. 
This is reflected in the smearing of the ADOS peaks. Despite the smearing, the remnants of high spectral weight around 
the FB energies still persist, implying that the FB physics remains relevant even in the weakly disordered regime. 
This robustness is essential, as in real systems, the presence of disorder is inevitable. The high density of states at 
the FB energies also enhances interaction effects, even at moderate coupling strengths, making the system a promising 
candidate for realizing correlated states, such as ferromagnetism or fractional Chern insulators in the presence of 
interactions and topology. Note that, as we increase the value of $\Delta$, the FB-induced peaks broaden 
more (see Fig.~\ref{fig:DOS}(c) with $\Delta = 0.5t$), and approximately around $\Delta = 0.8t$, the FB-induced peaks 
almost merge into the continuum (see Fig.~\ref{fig:DOS}(d)). In order to substantiate the results on the ADOS, we have 
also computed the inverse participation ratio (\textit{IPR}) for these various onsite disorder strengths. This is shown 
in Appendix~\ref{Appendix-B}.

\section{Characterization of the topological states}
\label{sec:topology}
\begin{figure*}[ht]
\centering
\includegraphics[clip, width=0.325\textwidth]{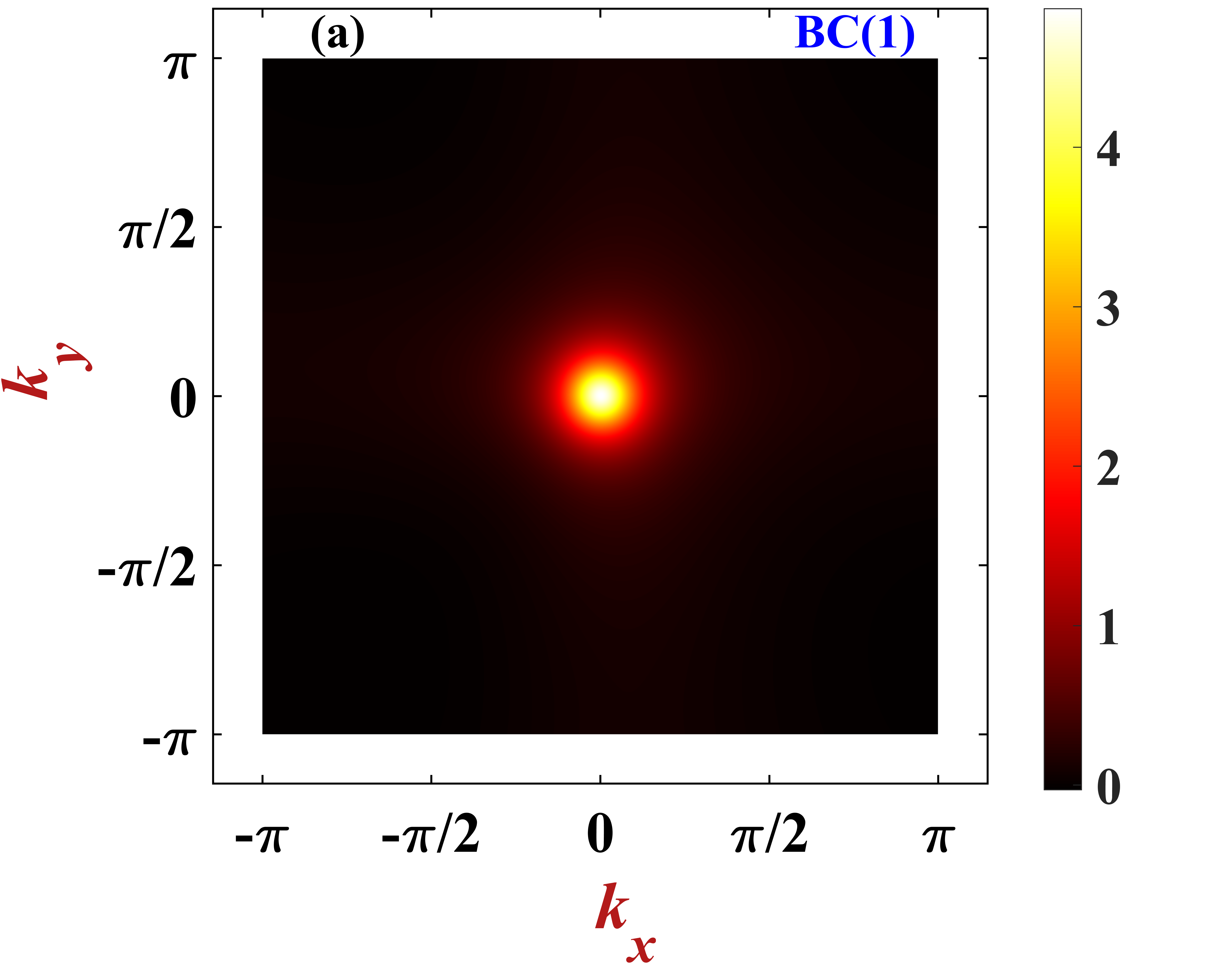}
\includegraphics[clip, width=0.325\textwidth]{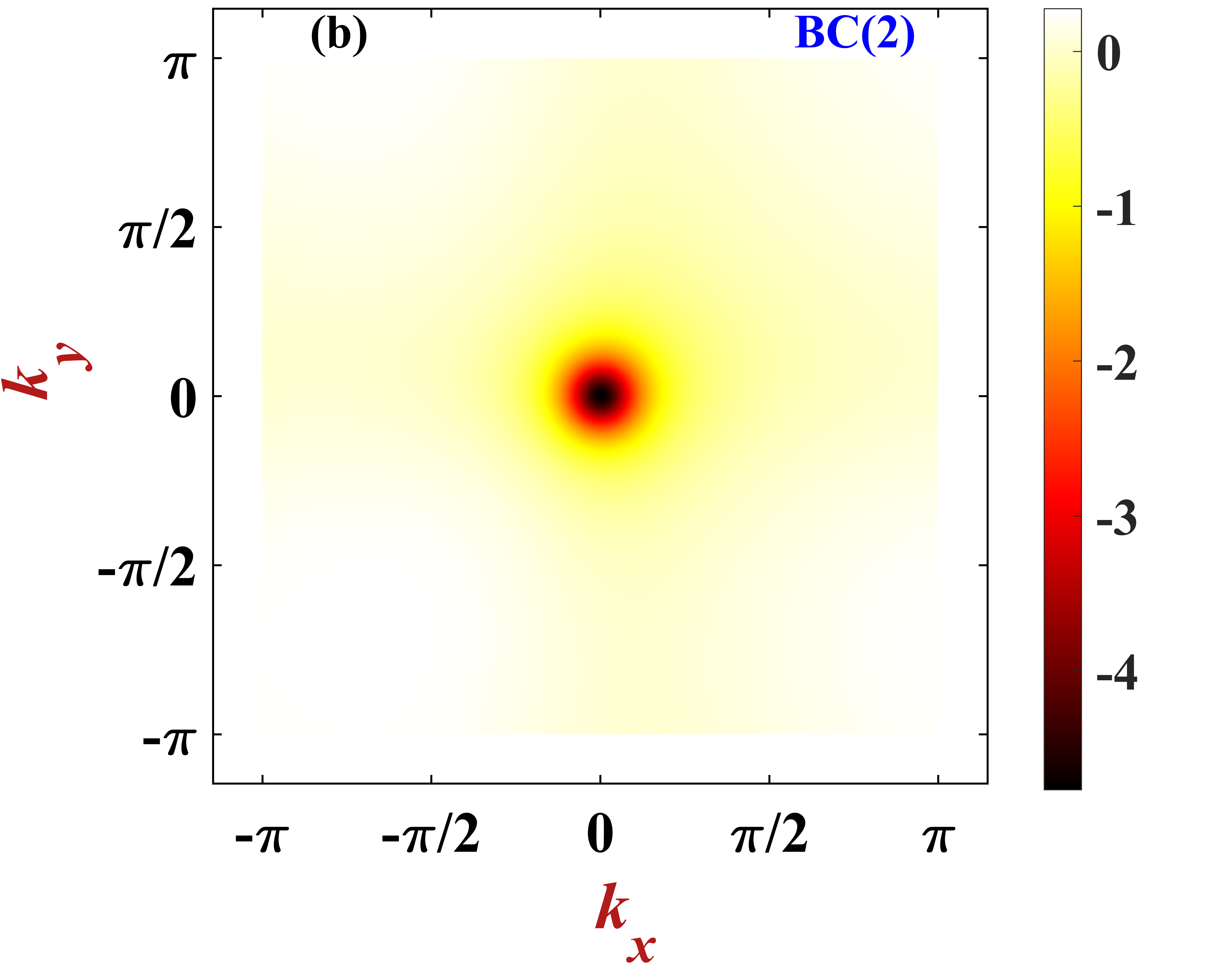}
\includegraphics[clip, width=0.325\textwidth]{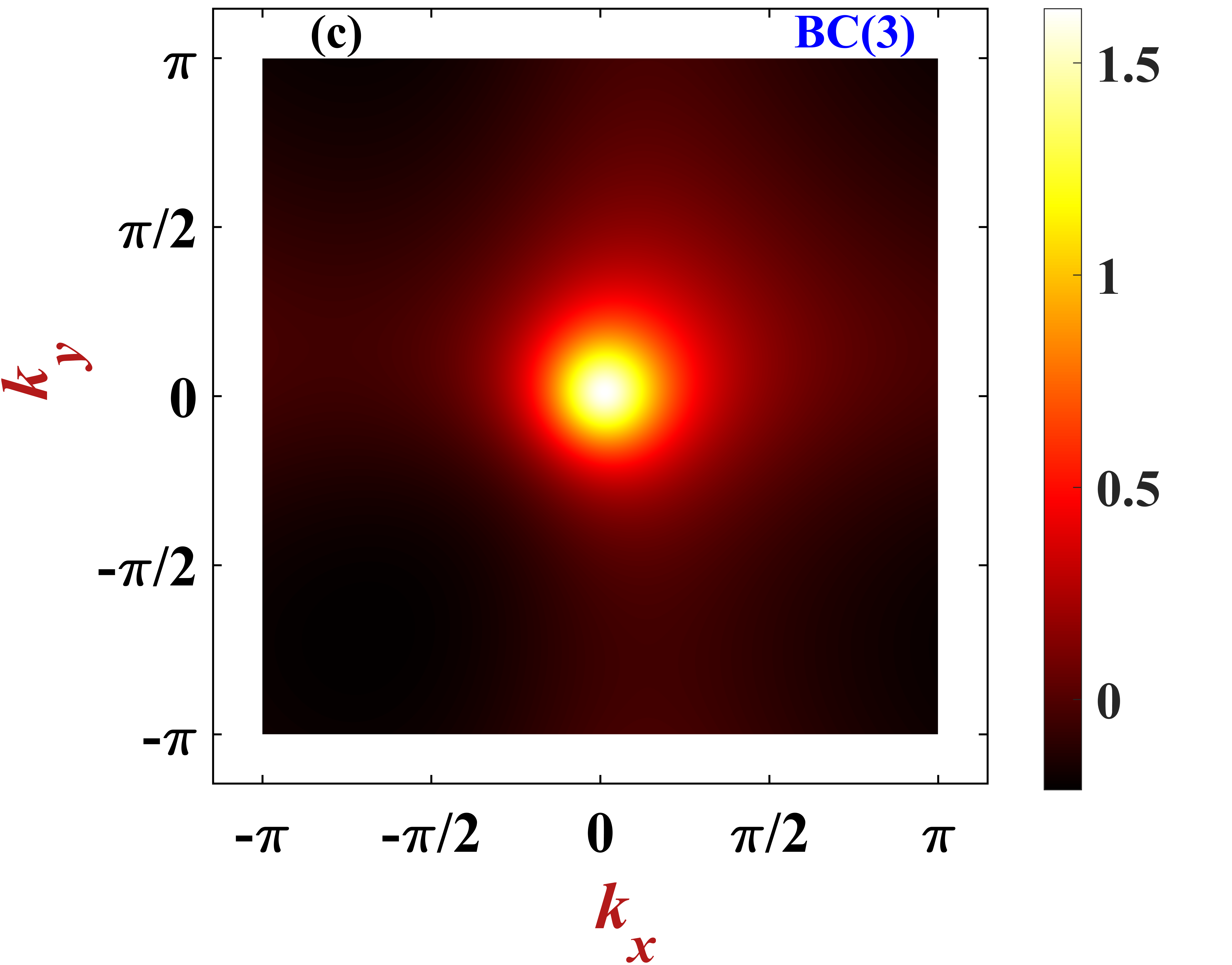} \\
\vskip 0.1cm
\includegraphics[clip, width=0.325\textwidth]{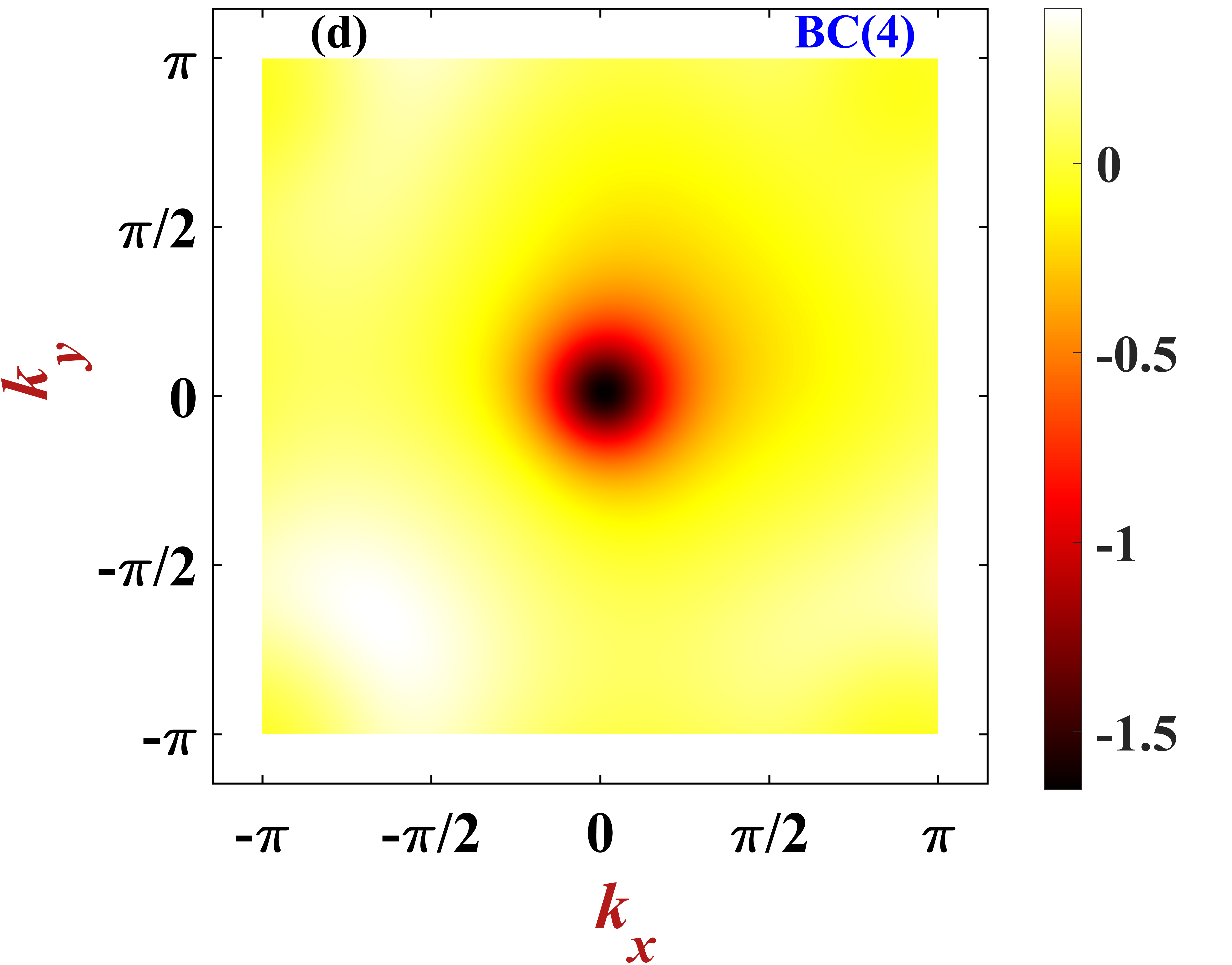}
\includegraphics[clip, width=0.325\textwidth]{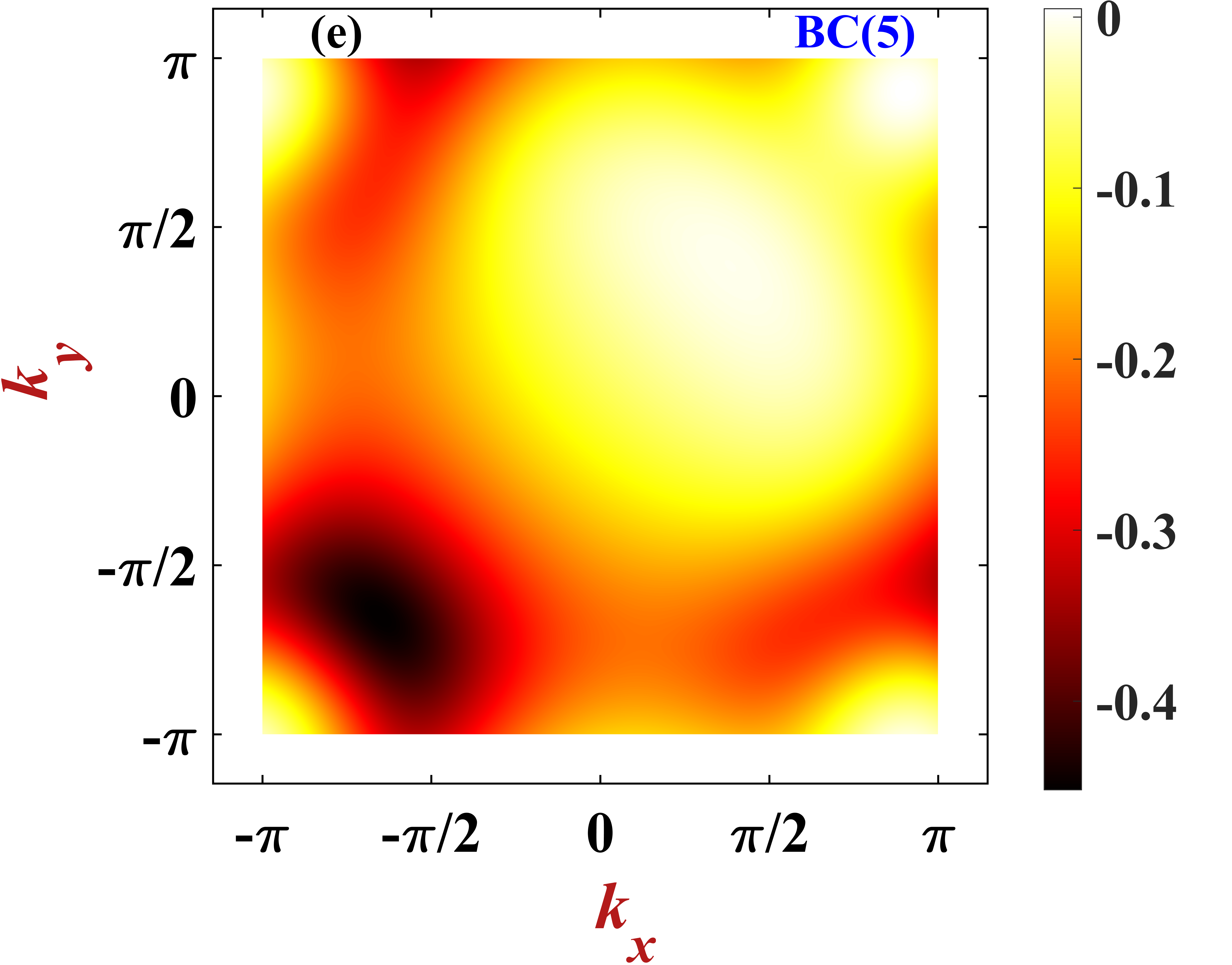}
\includegraphics[clip, width=0.325\textwidth]{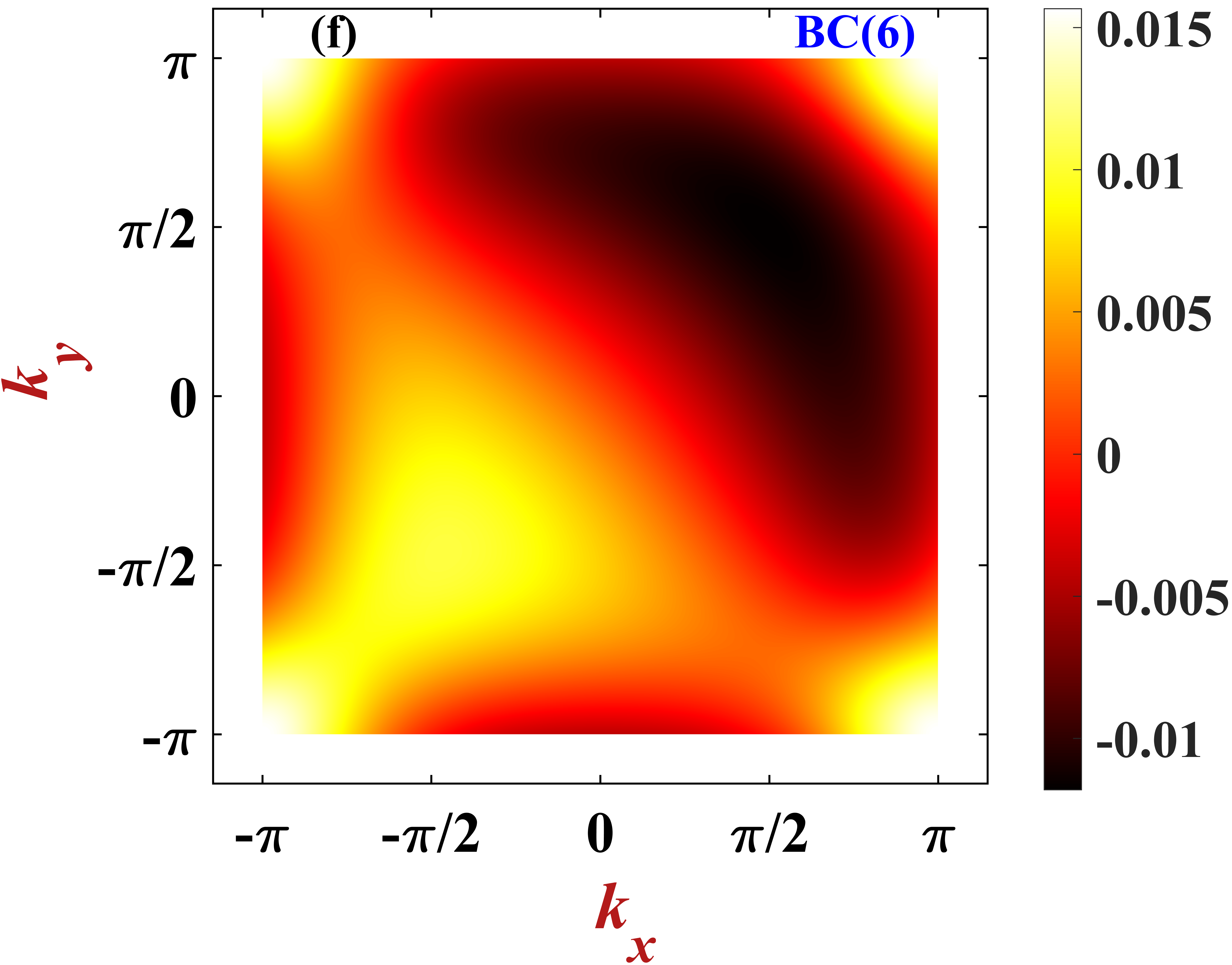}
\caption{Distribution of the Berry curvatures (BC) in the first Brillouin zone 
for all the six bands ($n = 1$ to $6$) in the diamond-dodecagon lattice at a 
magnetic flux $\Phi = \Phi_{0}/4$, corresponding to the band structure shown 
in Fig.~\ref{fig:band-structure}(b). The upper panel corresponds to the band 
index $n=1$ to $3$ and the lower panel corresponds to the band index $n=4$ to 
$6$. The system exhibits nontrivial topology with Chern numbers $C_{1} = +1$ 
and $C_{5} = -1$, while the remaining bands are topologically trivial 
($C_{n} = 0$ for $n=2,3,4,6$).}
\label{fig:Berry-curvature_Phi_0.25}
\end{figure*}

In this section, we will present the nontrivial topological properties of the diamond-dodecagon lattice model. Topological 
properties in a lattice model refers to the certain global, symmetry-protected features of its band structure, which are in 
general robust against the local perturbations, like small disorder, weak interaction etc. Topological properties of a lattice 
model can be perceived by studying its Bloch Hamiltonian $\mathcal{H}(\bm{k})$, as the topological properties emerge in the 
system when the Bloch eigenstates of the Hamiltonian $\mathcal{H}(\bm{k})$ cannot be smoothly defined over the entire 
Brillouin zone of the lattice. This can be quantified by calculating certain \emph{topological invariant} quantities like the 
Berry curvatures and Chern numbers corresponding to all the energy bands in the system~\cite{Haldane-prl-2004,Chen-jpcm-2012}. 
It is to be noted that, topology is a property of the eigenstates of the Hamiltonian of the system and not of its eigenvalues. 

The Berry curvature corresponding to each the band of the present lattice model can be computed using the following expressions: 
\begin{equation}
\Omega_{n}(\bm{k}) = 
\sum_{E_{m} (\neq E_{n})}\hspace{-0.5mm} 
\dfrac{-2\operatorname{\Im}\big( \mathcal{F}_{nm}^{x} \: \mathcal{F}_{mn}^{y} \big)} 
{\big[E_{n}(\bm{k})-E_{m}(\bm{k})\big]^2},
\label{eq:Berry-curvature}
\end{equation} 
%
\begin{align*}
\textrm{where} \quad
& \mathcal{F}_{nm}^{x} = \big\langle \psi_{n}(\bm{k}) \big| \dfrac{\partial \mathcal{H}(\bm{k})}{\partial k_{x}} 
\big| \psi_{m}(\bm{k}) \big\rangle \\
\textrm{and} \quad
& \mathcal{F}_{mn}^{y} = \big\langle \psi_{m}(\bm{k}) \big| \dfrac{\partial \mathcal{H}(\bm{k})}{\partial k_{y}} 
\big| \psi_{n}(\bm{k}) \big\rangle. 
\end{align*}
Here, $\big| \psi_{n}(\bm{k}) \big\rangle$ is the eigenstate of the Bloch Hamiltonian $\mathcal{H}(\bm{k})$ in 
Eq.~\eqref{eq:Bloch-matrix} with an eigenvalue $E_{n}(\bm{k})$, for the $n$-th band of the system. 
We can easily obtain the Chern number of the $n$-th band of the system by taking the summation over all the values of 
$\Omega_{n}(\bm{k})$ over the first Brillouin zone ($BZ$) of the lattice model, and is given by, 
\begin{equation}
C_{n} = \frac{1}{2 \pi}\int_{BZ}
\Omega_{n}(\bm{k})d{\bm{k}}.
\label{eq:Chern-number}
\end{equation}
It is apparent from the Eq.~\eqref{eq:Berry-curvature} that, the Berry curvature diverges if there is any band-touching 
between the energy bands of the system. Therefore, in order to evaluate the Berry curvatures and the Chern numbers for 
all the bands of the system, we need to open up the band-gap between all the energy bands in the system. For our model, 
we do so by tuning the nonzero values of the external magnetic flux $\Phi$ through the diamond plaquettes, which breaks the 
time-reversal symmetry of system and generates band-gaps between all the bands (cf. Fig.~\ref{fig:band-structure}(b) and (c)). 
\begin{figure*}[ht]
\centering
\includegraphics[clip, width=0.325\textwidth]{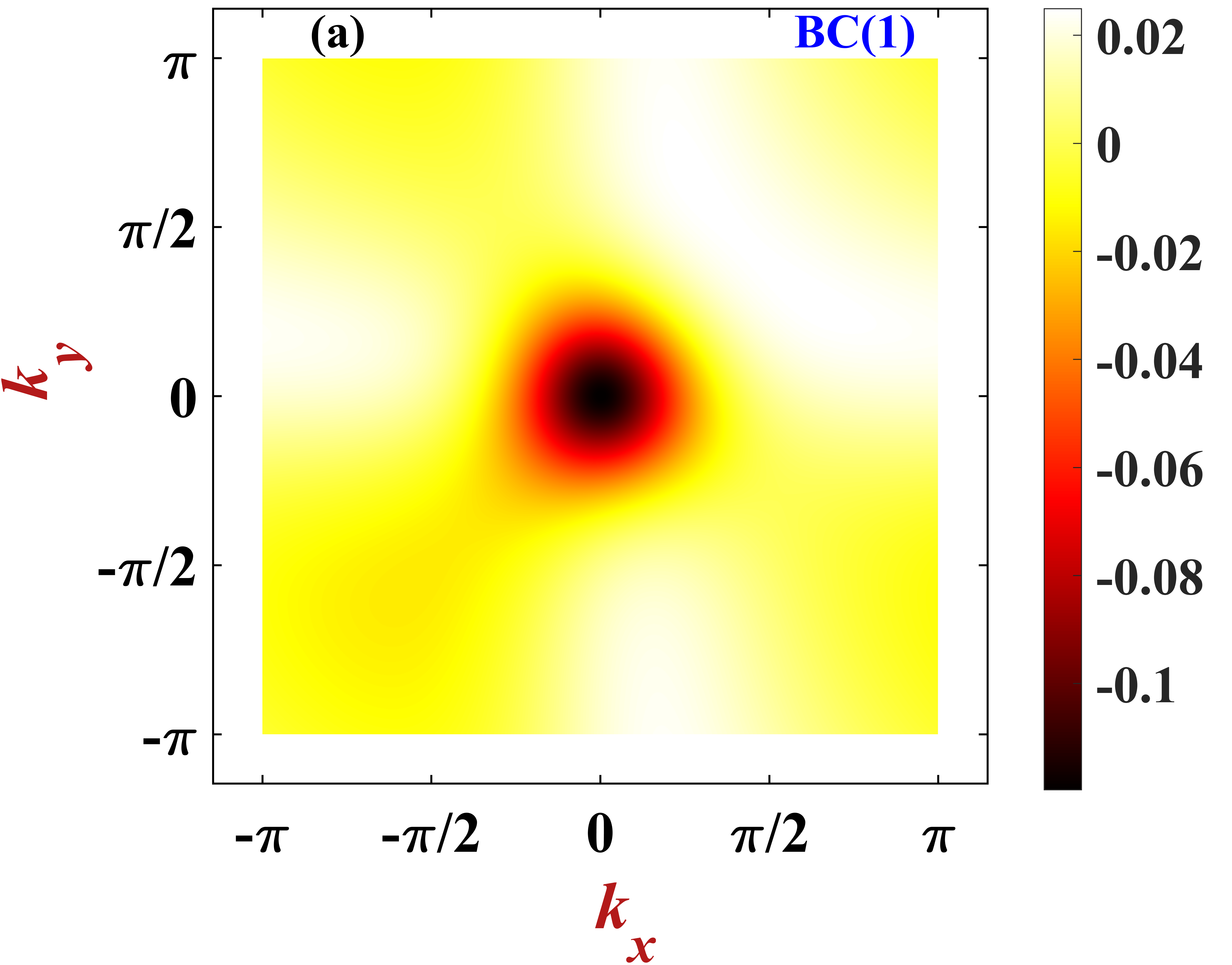}
\includegraphics[clip, width=0.325\textwidth]{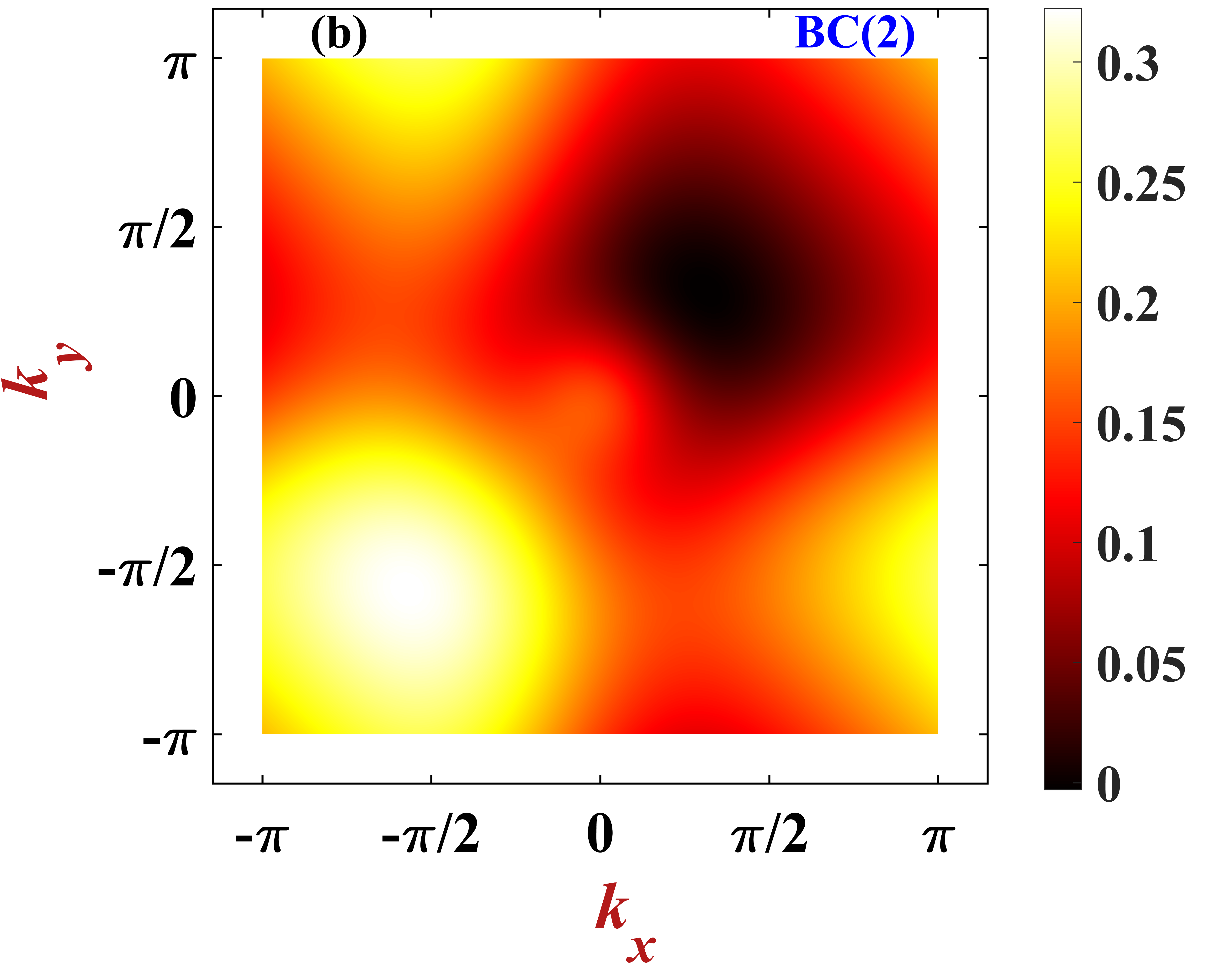}
\includegraphics[clip, width=0.325\textwidth]{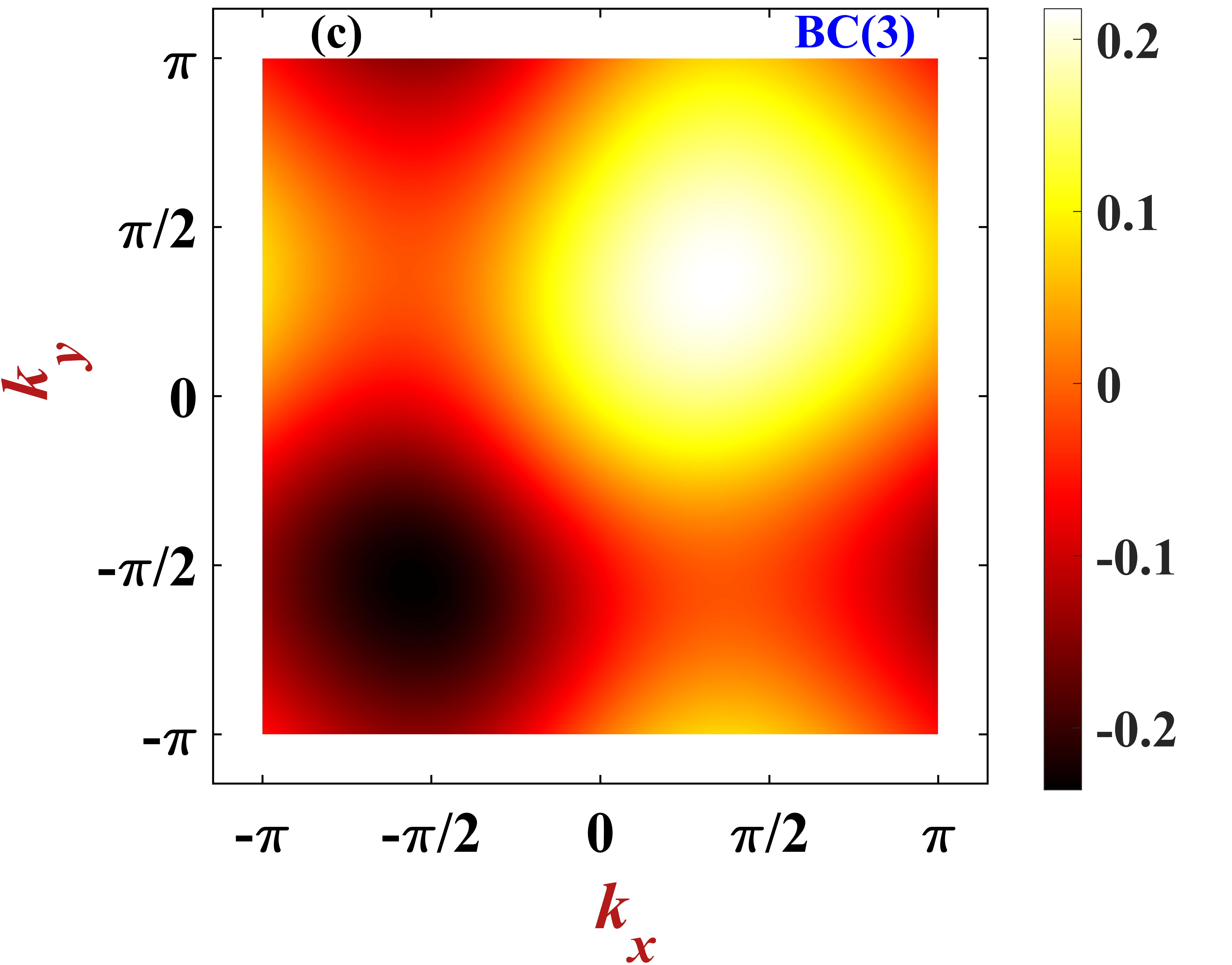} \\
\vskip 0.1cm
\includegraphics[clip, width=0.325\textwidth]{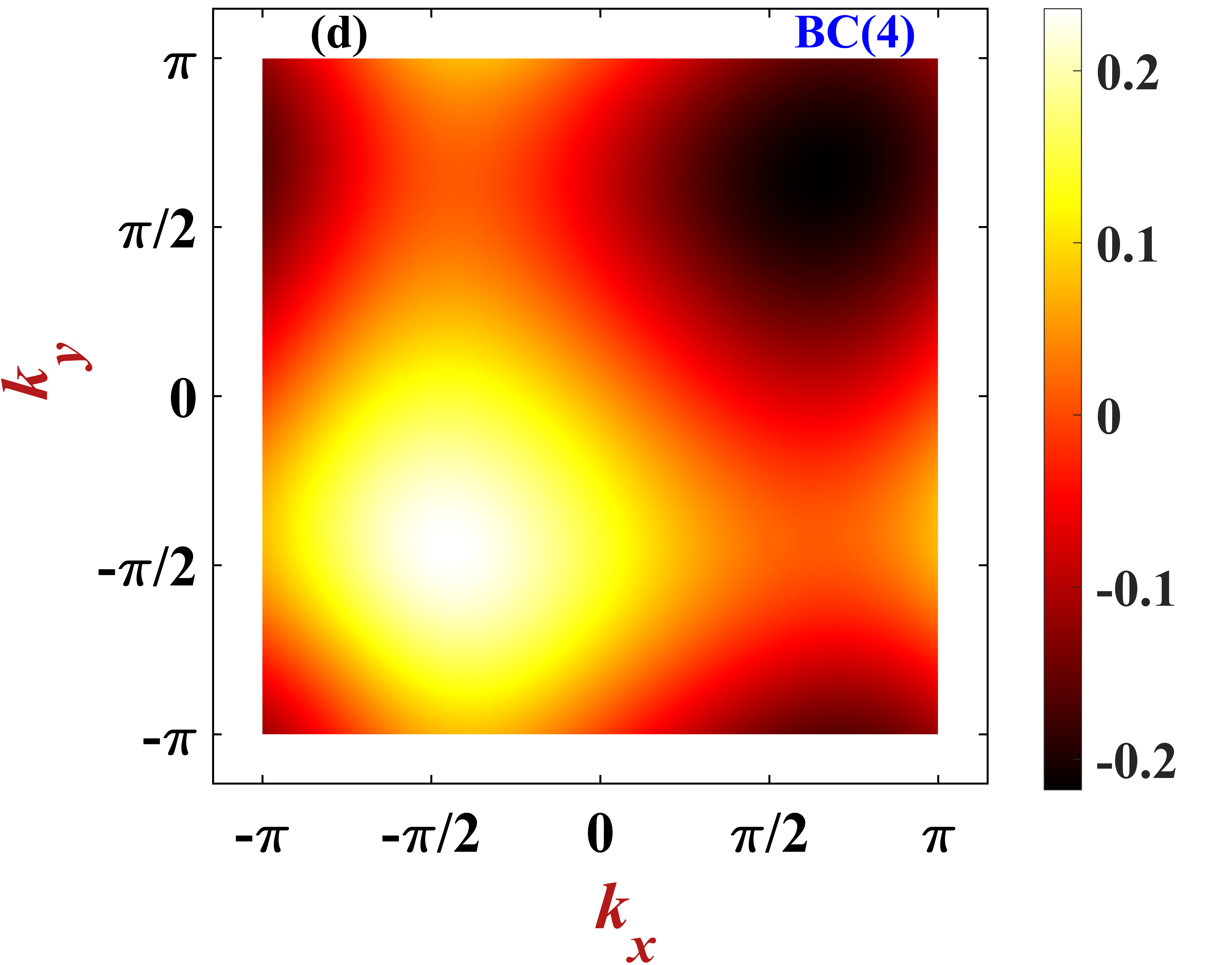}
\includegraphics[clip, width=0.325\textwidth]{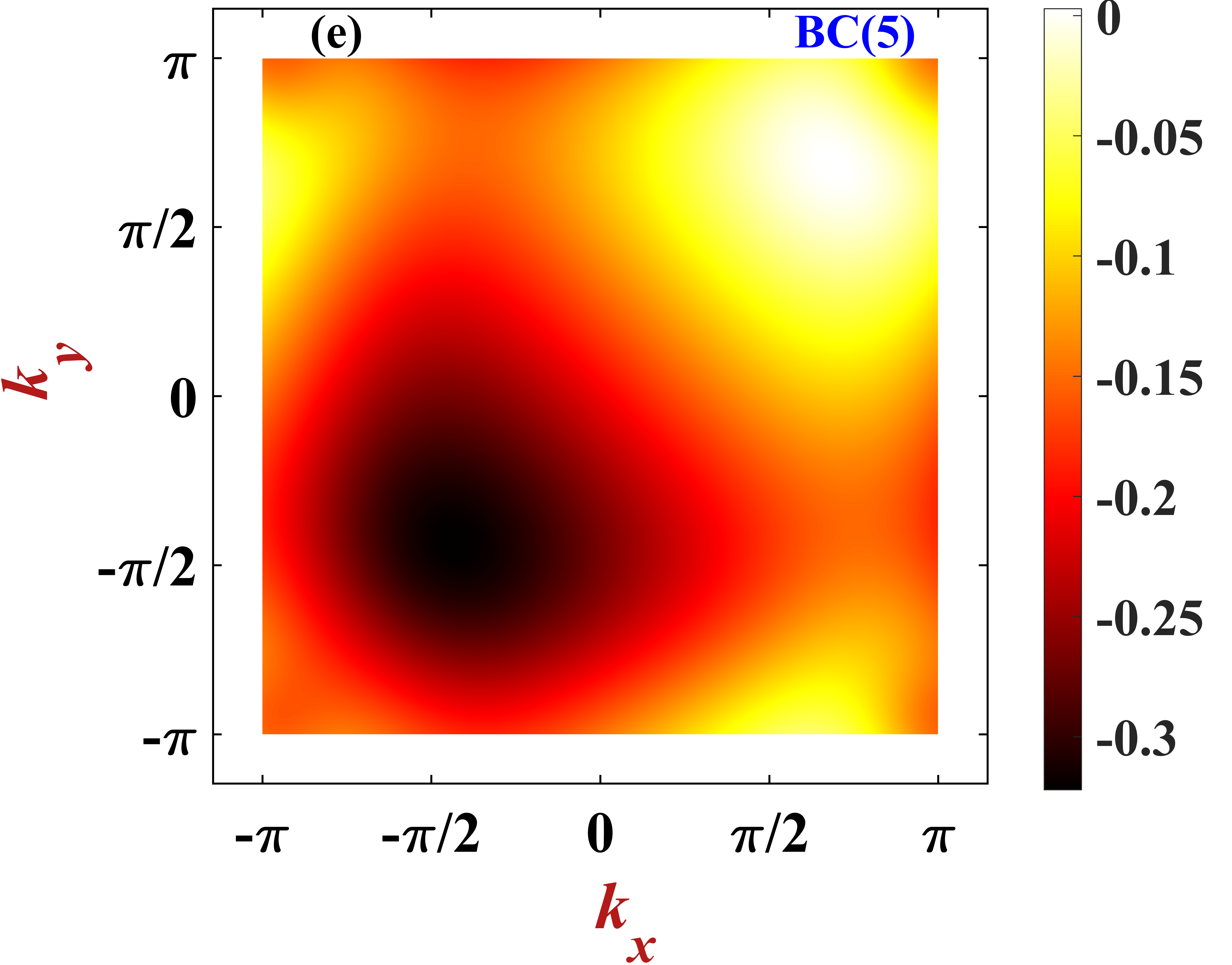}
\includegraphics[clip, width=0.325\textwidth]{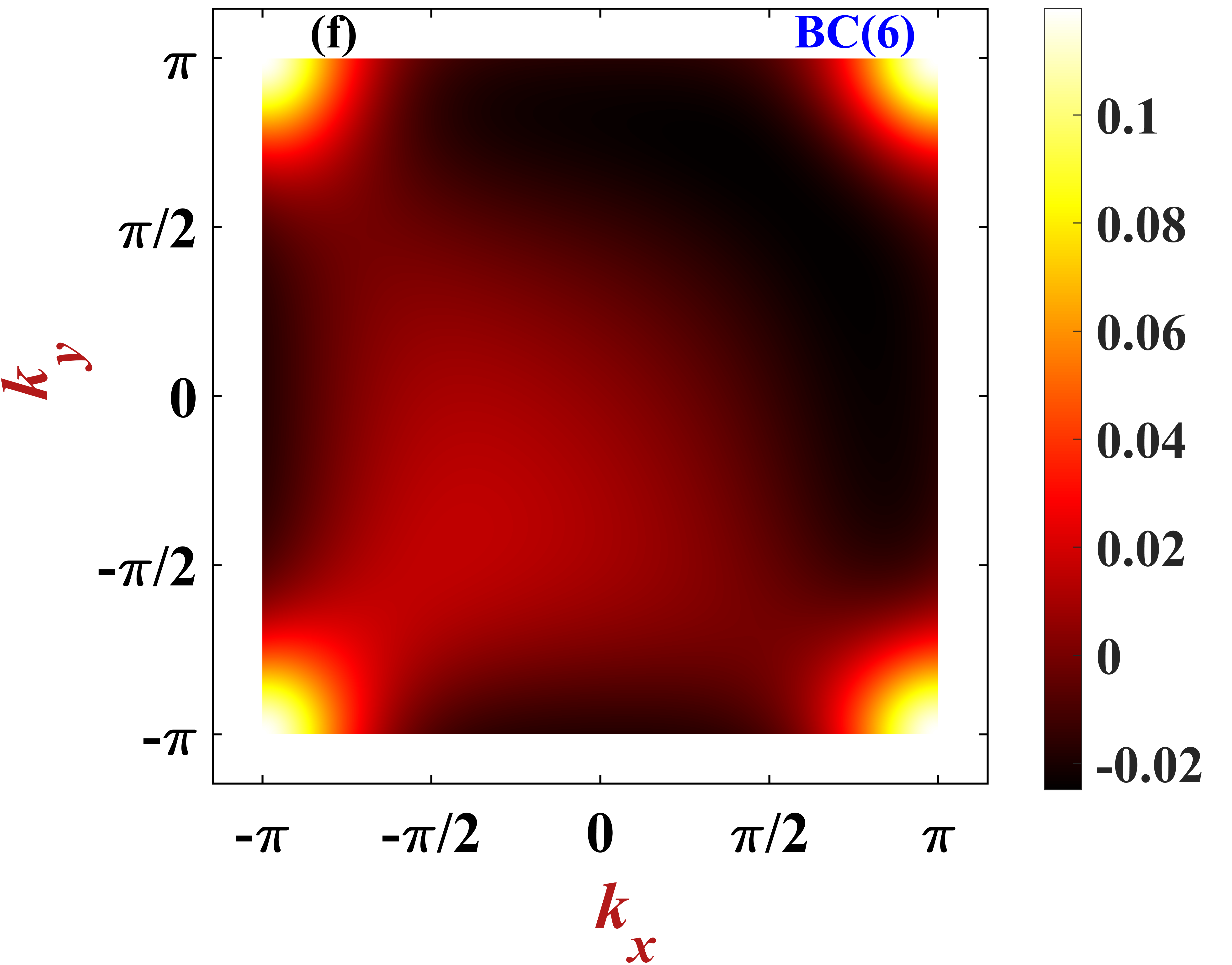}
\caption{Distribution of the Berry curvatures (BC) in the first Brillouin zone 
for all the six bands ($n = 1$ to $6$) in the diamond-dodecagon lattice at a 
magnetic flux $\Phi = \Phi_{0}/2$, corresponding to the band structure shown 
in Fig.~\ref{fig:band-structure}(c). The upper panel represents the plots for 
the band index $n=1$ to $3$ and the lower panel shows the plots for the band 
index $n=4$ to $6$. The second and the fifth bands exhibits nontrivial topological 
character with integer values of the Chern numbers $C_{2} = +1$ and $C_{5} = -1$, 
respectively, while the remaining bands are of topologically trivial in nature 
($C_{n} = 0$ for $n=1,3,4,6$).}
\label{fig:Berry-curvature_Phi_0.5}
\end{figure*}
%
\begin{figure*}[ht]
\centering
\includegraphics[clip, width=1.0\textwidth]{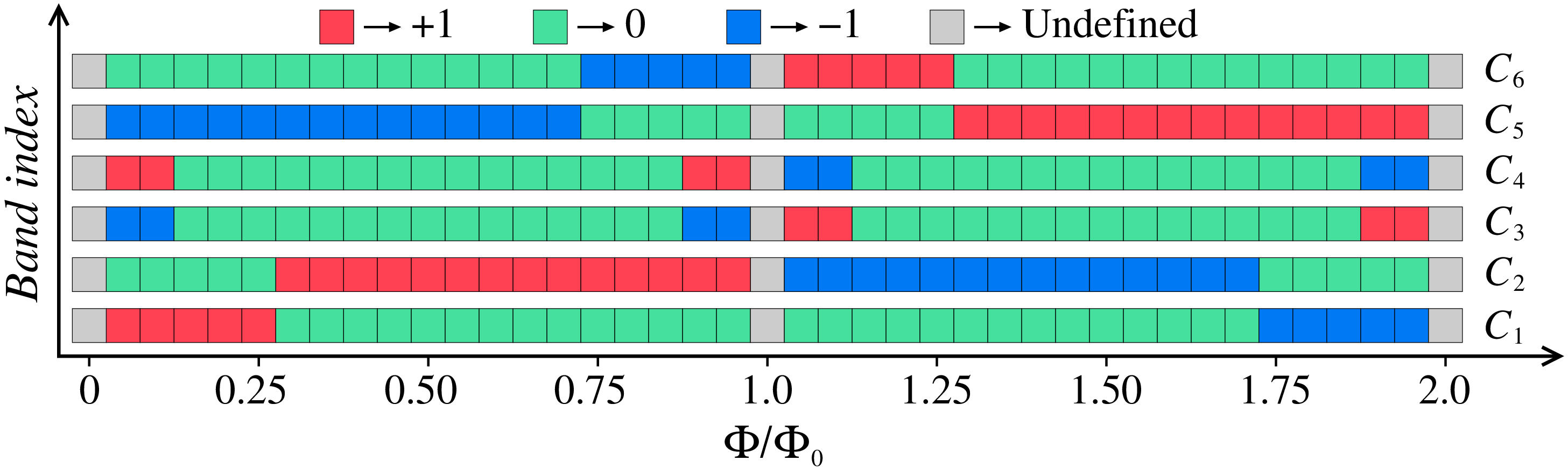}
\caption{Continuous evolution of the Chern numbers as a function of the 
magnetic flux $\Phi$ for all the six bands $n=1$ to $n=6$ in the energy spectrum. There 
are three different integer values of the Chern numbers, \textit{viz.}, $+1$ (red boxes), 
$0$ (green boxes), and $-1$ (blue boxes). The Chern numbers are not defined for $\Phi=0$, 
$\Phi_{0}$ and $2\Phi_{0}$, as some of the bands in the energy spectrum are touching with 
each other at those flux values.}
\label{fig:Chern-number}
\end{figure*}
%
%
\begin{figure}[ht]
\centering
\includegraphics[clip, width=1.0\columnwidth]{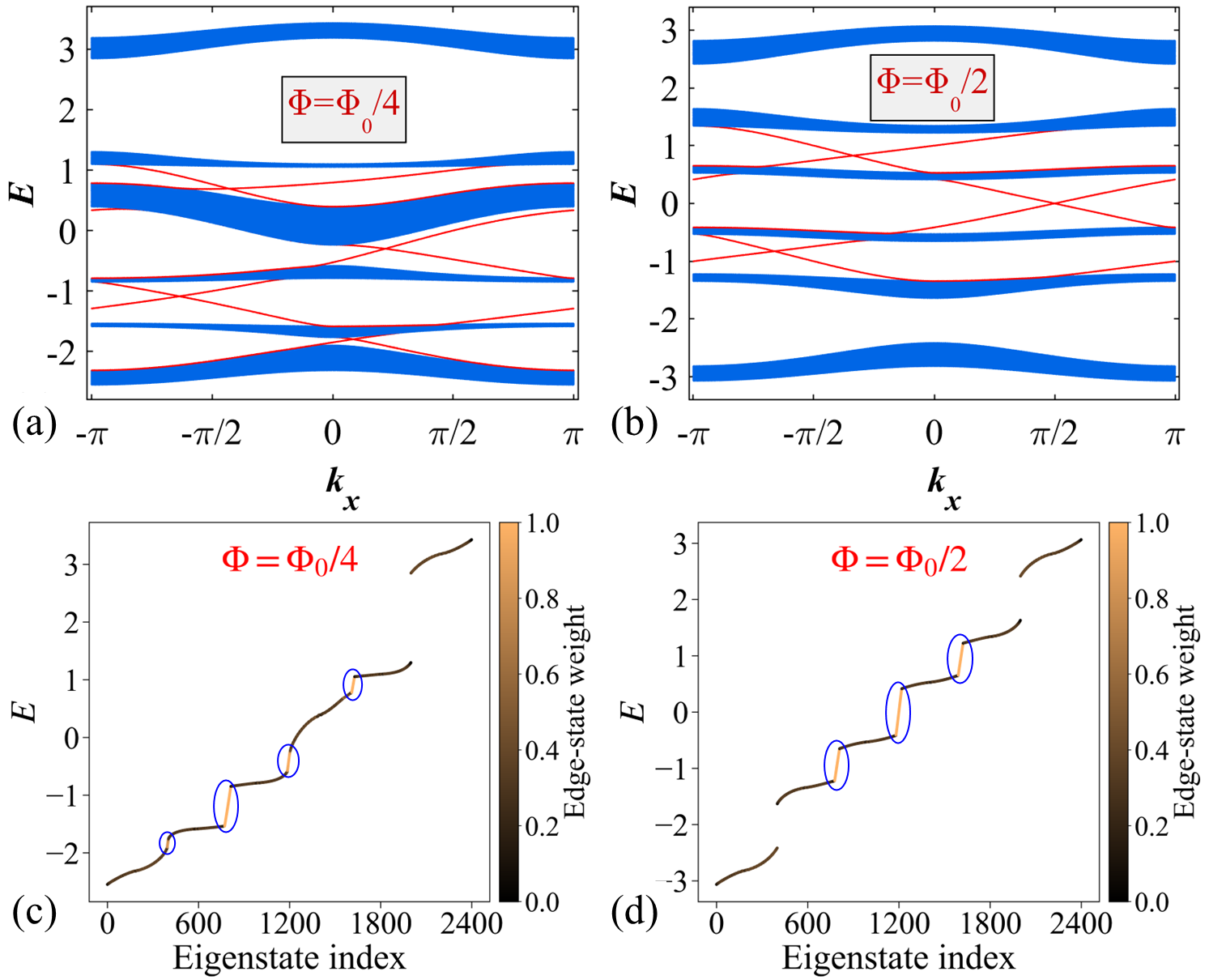}
\caption{Edge-state profiles for a finite-size diamond-dodecagon lattice model corresponding 
to the flux values (a) $\Phi=\Phi_{0}/4$ and (b) $\Phi=\Phi_{0}/2$. The light red lines correspond 
to the edge states while the dark blue bands represent to the bulk bands, consistent with the band 
structure diagrams in Figs.~\ref{fig:band-structure}(b) and (c). We have taken open boundary 
condition along the $y$-direction ($100$ chains are considered) and periodic boundary condition 
along the $x$-direction. We have encoded the same information also from the energy eigenvalues 
vs. eigenstate index plots for a $20 \times 20$ lattice system with (c) $\Phi=\Phi_{0}/4$ and 
(d) $\Phi=\Phi_{0}/2$ flux values. The light color represents the edge-states eigenvalues (marked with 
closed loops), while the dark color represents the bulk-states eigenvalues.}
\label{fig:Edge-states}
\end{figure}
%
\begin{figure*}[ht]
\centering
\includegraphics[clip, width=0.95\textwidth]{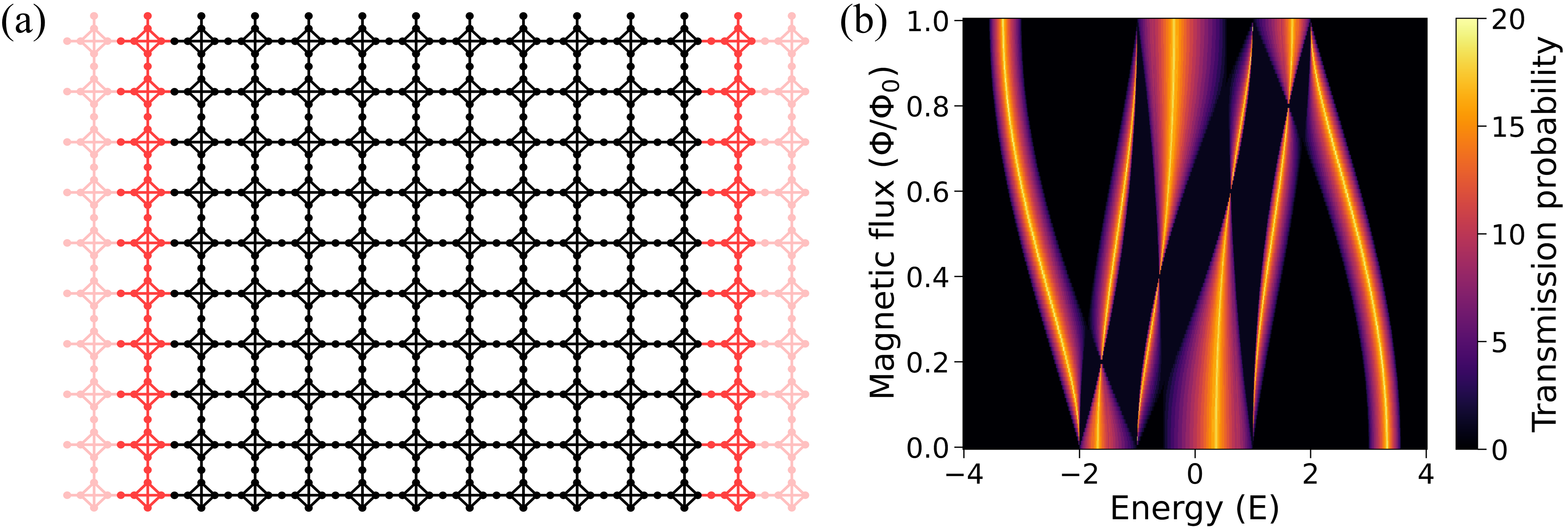}
\caption{(a) Schematic diagram of a finite-size diamond-dodecagon 
lattice scattering setup. The black dots and connecting lines denote the central 
scattering region, while the red sites and links indicate the semi-infinite leads 
attached along the $\pm x$ directions. 
(b) Density-plot of the two-terminal transmission probability $T(E, \Phi)$ through 
the finite-size $20 \times 20$ diamond-dodecagon lattice system as a function of 
magnetic flux $\Phi$ measured in units of $\Phi_{0}$ (vertical axis) and the energy 
of the incoming electrons $E$ (horizontal axis). The color scale (shown at the right) 
encodes $T$ from 0 (dark) to its maximum (bright). The computation is done at zero 
temperature and we set $t=\lambda=1.0$ eV.}
\label{fig:transport}
\end{figure*}
 
Under this condition, we have computed the Berry curvatures and hence the Chern numbers for all the isolated bands of the 
system to understand the topological character of the Bloch eigenstates in this lattice model. We have chosen two different 
nonzero values of the magnetic flux: $\Phi=\Phi_{0}/4$ and $\Phi=\Phi_{0}/2$ to obtain the numerical results for the 
topological invariants of the system. For the case of $\Phi=\Phi_{0}/4$, the numerical plots of the Berry curvature 
distributions in the $\bm{k}$-space (first Brillouin zone) corresponding to all the six bands ($n = 1$ to $6$) are 
displayed in Fig.~\ref{fig:Berry-curvature_Phi_0.25}(a)-(f). The corresponding values of the Chern numbers are 
$C_{n} = \{+1,0,0,0,-1,0\}$. Thus, it is clearly evident that, the lowest band ($n=1$) and the fifth band ($n=5$) 
exhibit nontrivial topological character with nonzero integer values of the Chern numbers equal to $C_{1} = +1$ and 
$C_{5} = -1$, respectively. We have also computed the Berry curvature distributions for all the bands in the system for 
the flux $\Phi=\Phi_{0}/2$, which are shown in Fig.~\ref{fig:Berry-curvature_Phi_0.5}(a)-(f). In this case, the corresponding 
values of the Chern numbers for the different bands are $C_{n} = \{0,+1,0,0,-1,0\}$. Interestingly, now we observe that, the 
second band ($n=2$) and the fifth band ($n=5$) show nonzero integer values of the Chern numbers, $C_{2} = +1$ and $C_{5} = -1$, 
respectively, exhibiting their topological character. We remark that, it indicates towards a topological phase transition 
for this lattice model. We have also computed the Berry curvature distributions and the corresponding values of the Chern 
numbers for this lattice model with various other nonzero values of the flux starting from $\Phi=0$ to $\Phi=2\Phi_{0}$. This 
full phase diagram for the variation of the Chern numbers as a function of the magnetic flux corresponding to all the six 
bands $n=1$ to $n=6$ in the energy spectrum is shown in Fig.~\ref{fig:Chern-number}. It is clearly visible from the 
Fig.~\ref{fig:Chern-number} that, there is a continuous evolution of the Chern numbers corresponding to all the six bands as a 
function of the magnetic flux $\Phi$. Whenever there is a transition in the values of the Chern numbers, there is always a 
band gap closing and reopening between the two bands at those respective values of the magnetic flux. This has been cross-checked 
from the band structure analysis. Note that, we have obtained these results numerically using the Eq.~\eqref{eq:Berry-curvature} 
and Eq.~\eqref{eq:Chern-number}, and the $k$-resolution for the first Brillouin zone is taken to be $\Delta k = 0.001\pi$. 
In Fig.~\ref{fig:Edge-states}, we have shown the edge-state profiles corresponding to the topological phases found in 
cases of Figs.~\ref{fig:band-structure}(b) and (c). One can clearly see the bulk-boundary correspondence from Fig.~\ref{fig:Edge-states}, 
which is consistent with the information about the Chern numbers obtained from the Fig.~\ref{fig:Berry-curvature_Phi_0.25} and 
Fig.~\ref{fig:Berry-curvature_Phi_0.5}.

Before we end this section, we would like to emphasize on the fact that, such 2D lattice models exhibiting nearly flat 
topological bands with nonzero integer values of the Chern numbers are very useful for the realization of the quantum Hall 
states in lattice settings, which has been exhibited earlier in other interesting 2D lattice models also~\cite{Wen-prl-2011,
Das-Sarma-prl-2011,Neupert-prl-2011}. 

\section{Transport properties in a finite-size system}
\label{sec:transport}
In this section, we explore the transport properties of a finite-size diamond-dodecagon lattice setup as a function of the 
externally tunable magnetic flux $\Phi$ and the energy ($E$) of the incoming electrons. For this computation, we consider 
a finite segment of the 2D diamond-dodecagon lattice, composed of $20 \times 20$ arrays of unit cells 
(having $1200$ atomic sites), attached in-between two 2D semi-infinite leads along the $\pm x$ directions (see 
Fig.~\ref{fig:transport}(a)). The central scattering region (black sites and bonds) is bounded by two identical leads (red sites 
and bonds), modelled as the perfect repetitions of the unit cell geometry, enabling coherent quantum transport along the longitudinal 
direction. 

The transport characteristics are calculated using the \textit{scattering matrix formalism}~\cite{Datta1,Datta2}, as 
implemented in the \texttt{Kwant} package~\cite{Groth2014}, which enables the efficient computation of the transmission 
probabilities for any arbitrary tight-binding lattice geometries connected to external leads~\cite{Majhi2022,Istas2019,
Waintal2024,Li2018}. Mathematically, the scattering matrix $S$ relates the incoming and the outgoing modes in the leads, 
and can be used to extract the total transmission probability between any two terminals. This method is formally equivalent 
to the non-equilibrium Green's function (NEGF) technique, as established by the Fisher-Lee relation~\cite{Fisher1981}, 
where the self-energy terms from the semi-infinite leads account for their influence on the central region as plane-wave 
superpositions. 

The total transmission probability from lead $p$ (known as the \textit{source}) to lead $q$ (known as the \textit{drain}) at 
an energy $E$ is computed as~\cite{Groth2014}: 
\begin{equation}
T(E) = \sum_{a \in p} \sum_{b \in q} \big| S_{pq}^{(ba)}(E) \big|^2, 
\label{eq:transmission}
\end{equation}
where $S_{pq}^{(ba)}$ is the scattering matrix element describing the amplitude of a wave transmitted from mode $a$ in the 
lead $p$ to mode $b$ in the lead $q$. 
To obtain the scattering matrix $S$, the time-independent Schr\"{o}dinger equation
\begin{equation}
\bm{H}_{\textrm{full}} \psi_{i} = E \psi_{i}
\end{equation}
is solved, where $\bm{H}_{\textrm{full}}$ is the full Hamiltonian of the system written in a block-tridiagonal form that 
includes the Hamiltonian of the scattering region, the coupling matrices between the scattering region and the leads, and 
the unit cell Hamiltonian of the leads. The semi-infinite nature of the leads is incorporated via the self-energy terms, 
which project the semi-infinite lead degrees of freedom onto the boundary sites of the scattering region through their 
surface Green's functions. The wave function $\psi_i$ describes the state of the system inside the scattering region, 
matched to the propagating and evanescent modes in the leads.

Using the scattering matrix approach outlined in Eq.~\eqref{eq:transmission}, we have computed the transmission characteristics 
for the system depicted in Fig.~\ref{fig:transport}(a) as a function of the magnetic flux $\Phi$ and the energy ($E$) of the 
electron as shown in Fig.~\ref{fig:transport}(b). It shows an interesting and rich phase diagram, exhibiting regions of complete 
transmission suppression [dark spots in Fig.~\ref{fig:transport}(b)] coexisting with regions of ballistic transmission [bright 
spots in Fig.~\ref{fig:transport}(b)], and also re-entrant transitions in between them as a function of the magnetic flux $\Phi$ 
and the energy ($E$) of the electron. It unfolds an interesting possibility of opening and closing the conducting channels in this 
system by tuning the externally controllable parameter $\Phi$. In addition to this, the complete suppression of the transmission 
probabilities corresponding to the FB energies is also indicative of the fact that they form the CLS and hence do not contribute 
to the transport. We note that, the physical origin of the resonant transmission peaks are the contributions 
coming from the dispersive bands. 

This analysis highlights the diamond-dodecagon lattice as a tunable quantum platform, where both FB-induced localization and 
topological effects can be investigated via the transport signatures of the system. The ability to modulate the transmission 
spectrum with an external magnetic flux suggests potential applications of this lattice model in programmable quantum devices, 
synthetic gauge field simulations, and topological insulator analogs in photonic or ultracold atom settings. 

\section{Summary and Future Outlook}
\label{sec:conclu}
In this work, we have presented a comprehensive theoretical study of a novel 2D tight-binding diamond-dodecagon lattice 
model, exploring its various intriguing features in the context of flat-band physics, topological properties and transport 
characteristics. We have shown that an interplay between the lattice geometry, externally tunable magnetic flux, and the 
quantum interference effect brings out these interesting properties in this lattice model. It has been shown that this 
lattice model exhibits multiple completely flat bands in its band structure, supported by exact constructions of the 
corresponding compact localized states in the real space. The computation of the density of states accomplishes that, 
these flat bands are very robust in their character in the presence of a weak disorder in the system. 
%
\renewcommand\thefigure{A.\arabic{figure}}    
\setcounter{figure}{0} 
\begin{figure*}[ht]
\centering
\includegraphics[clip, width=0.95\textwidth]{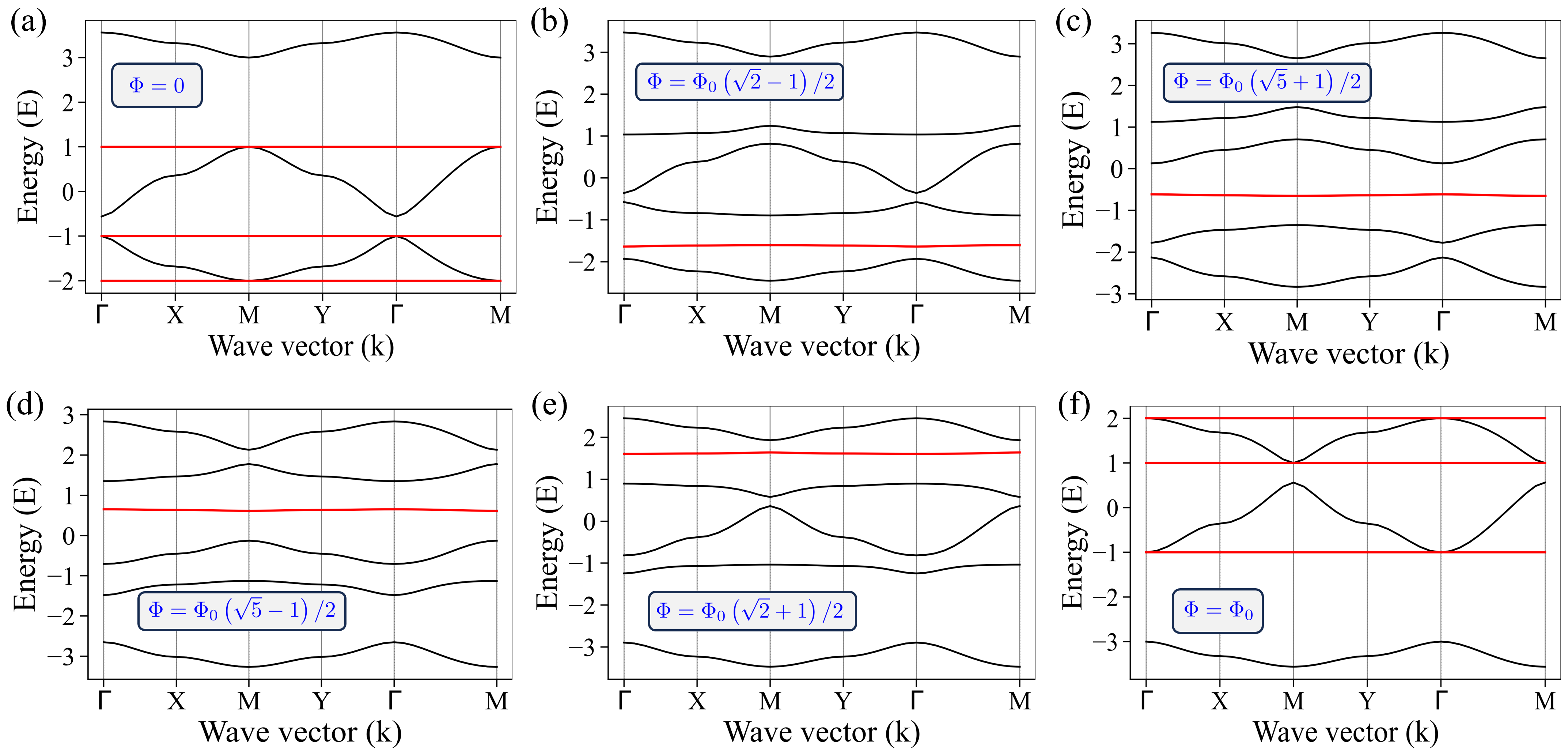}
\caption{Variation of the electronic band structure in the diamond-dodecagon 
lattice as a function of magnetic flux $\Phi$. The six figures show the energy 
dispersion along high-symmetry directions in the Brillouin zone for 
six different flux values: (a) $\Phi = 0$, (b) $\Phi = \Phi_0(\sqrt{2} - 1)/2$, 
(c) $\Phi = \Phi_0(\sqrt{5} + 1)/2$, (d) $\Phi = \Phi_0(\sqrt{5} - 1)/2$, 
(e) $\Phi = \Phi_0(\sqrt{2} + 1)/2$, and (f) $\Phi = \Phi_0$, where $\Phi_0$ is 
the fundamental flux quantum. The red horizontal lines indicate the perfectly 
flat bands with zero group velocity, while the black curves represent the 
dispersive bands. It is clearly visible from these plots that we can make any 
of the six bands completely flat at will by tuning certain specific values of 
the flux $\Phi$. Here the high‑symmetry points are $\Gamma$: $(0,0)$, X: $(\dfrac{\pi}{a},0)$, 
M: $(\dfrac{\pi}{a},\dfrac{\pi}{a})$, and Y: $(0,\dfrac{\pi}{a})$.}
\label{fig:FB_at_will}
\end{figure*}

Under the application of an external magnetic flux, the band gap opens up between all the bands in the system, making these 
completely flat bands into quasi-flat bands with nontrivial topological features. These nontrivial topological features are 
marked by the computation of the Berry curvatures and the Chern numbers for those topologically nontrivial bands. This opens up 
an interesting platform for exploring the quantum Hall physics and Chern insulator in this lattice model, both without and 
with interaction. The transmission characteristics evaluated for this system using the scattering matrix formalism also 
exhibit a rich pattern with multiple re-entrant transitions between the completely ballistic and the zero transmission 
domains as a function of the externally controllable magnetic flux. This indicates an interesting possibility of utilizing 
this system for potential applications in quantum sensor devices. Additionally, flux tunability of the spectrum offers dynamic 
control over transport, essential for exploring flux-induced topological transitions and magnetotransport responses.  

The proposed diamond-dodecagon lattice model offers fertile ground for various intriguing future research directions. 
The ability to systematically control both the flatness and topological properties of multiple bands makes the 
diamond-dodecagon lattice particularly attractive for exploring various strongly correlated topological phenomena. 
Additionally, incorporating spin-orbit coupling in this model could reveal richer topological phases, including the 
time-reversal invariant topological insulators.  Furthermore, periodic driving (Floquet engineering) may dynamically 
modulate the band topology, opening avenues to explore the non-equilibrium topological states. From the experimental 
perspective, the diamond-dodecagon lattice can be realized in real-life experiments in the field of photonics using 
the laser-induced single-mode photonic waveguide arrays to explore the corresponding topological photonics phenomena.

\begin{acknowledgments}
BP would like to acknowledge the funding from the Nagaland University 
through a minor start-up research grant. 
\end{acknowledgments}
%
\appendix
\section{Magnetic flux-driven evolution of FBs}
\label{Appendix-A}

In this appendix, we present a systematic evolution of the electronic band structure in the diamond-dodecagon 
lattice as a function of the magnetic flux $\Phi$. We would like to point out a very interesting aspect 
observed in the band structure of this lattice model: One can systematically make any of the six bands in the 
band structure of this lattice model perfectly flat at will by tuning the value of the external magnetic flux 
$\Phi$. This interesting feature is highlighted in Fig.~\ref{fig:FB_at_will}. We remark that the energy 
dispersions in Fig.~\ref{fig:FB_at_will} are plotted along the high-symmetry directions in the Brillouin zone. 

As discussed previously, at $\Phi=0$, we find that the lowest band of the system exhibits perfect flatness 
along with two other complete flat bands also (see Fig.~\ref{fig:FB_at_will}(a)). As we tune the value of 
$\Phi = \Phi_0(\sqrt{2} - 1)/2$, we observe that the second band of the system becomes completely flat, while 
for $\Phi = \Phi_0(\sqrt{5} + 1)/2$, the third band of the system turns into a complete flat band (see 
Fig.~\ref{fig:FB_at_will}(b) and (c)). Similarly, for $\Phi = \Phi_0(\sqrt{5} - 1)/2$ and 
$\Phi = \Phi_0(\sqrt{2} + 1)/2$, the fourth and the fifth bands of the system, respectively, exhibit perfect 
flatness (see Fig.~\ref{fig:FB_at_will}(d) and (e)). Finally for $\Phi = \Phi_0$, the topmost band of the 
system transforms into a completely flat band along with two other perfectly flat bands in the spectrum 
(see Fig.~\ref{fig:FB_at_will}(f)). This interesting feature of controlling the position of the completely 
flat band in the band structure of the system at will offers us a very unique situation, where one can set the 
Fermi energy of the system to be aligned with flat band energy and study various strongly correlated phenomena 
for this lattice model at different filling factors.
\section{IPR with various onsite disorder}
\label{Appendix-B}
%
\renewcommand\thefigure{B.\arabic{figure}}    
\setcounter{figure}{0} 
\begin{figure}[ht]
\centering
\includegraphics[clip, width=0.49\columnwidth]{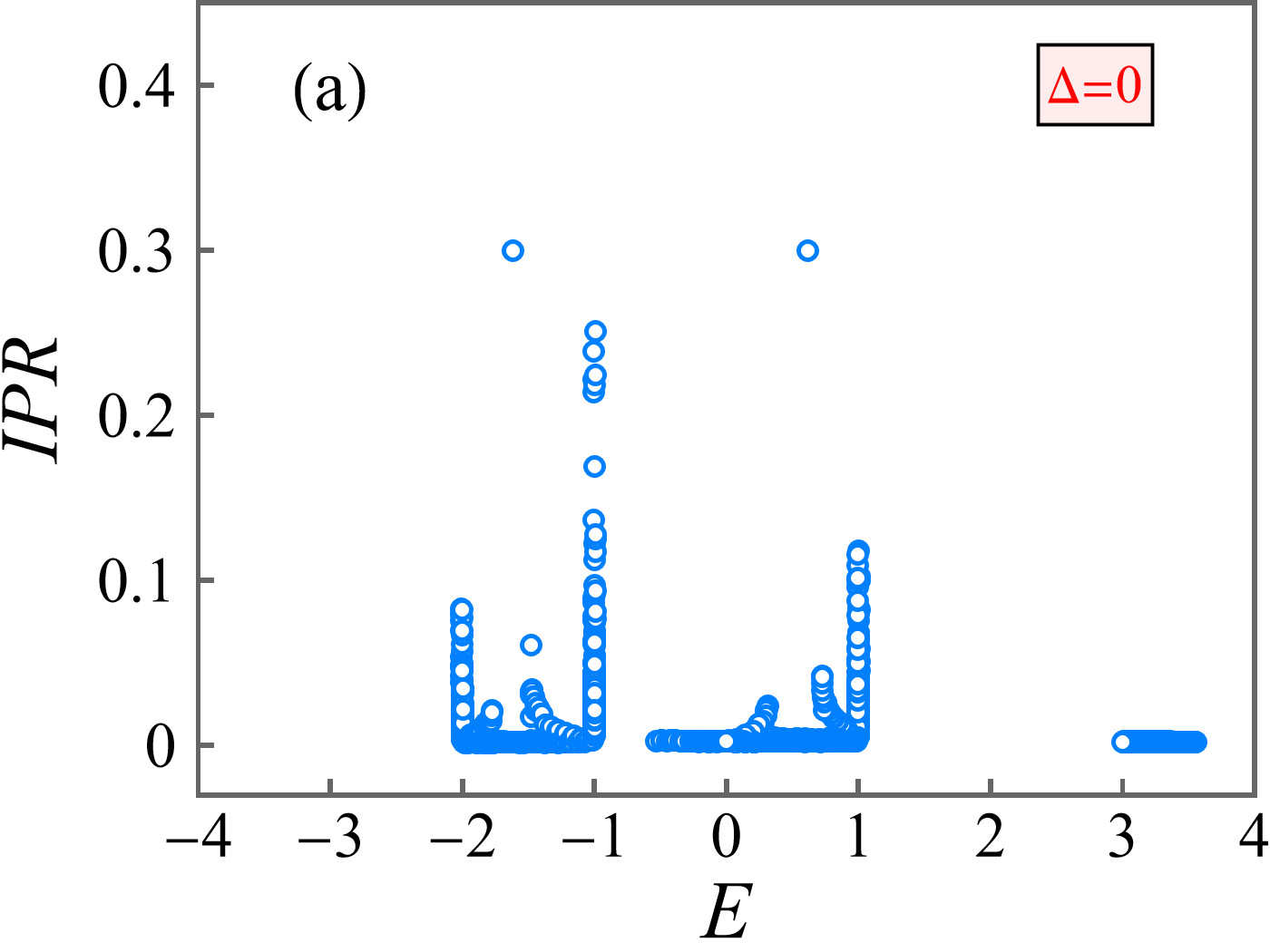}
\includegraphics[clip, width=0.49\columnwidth]{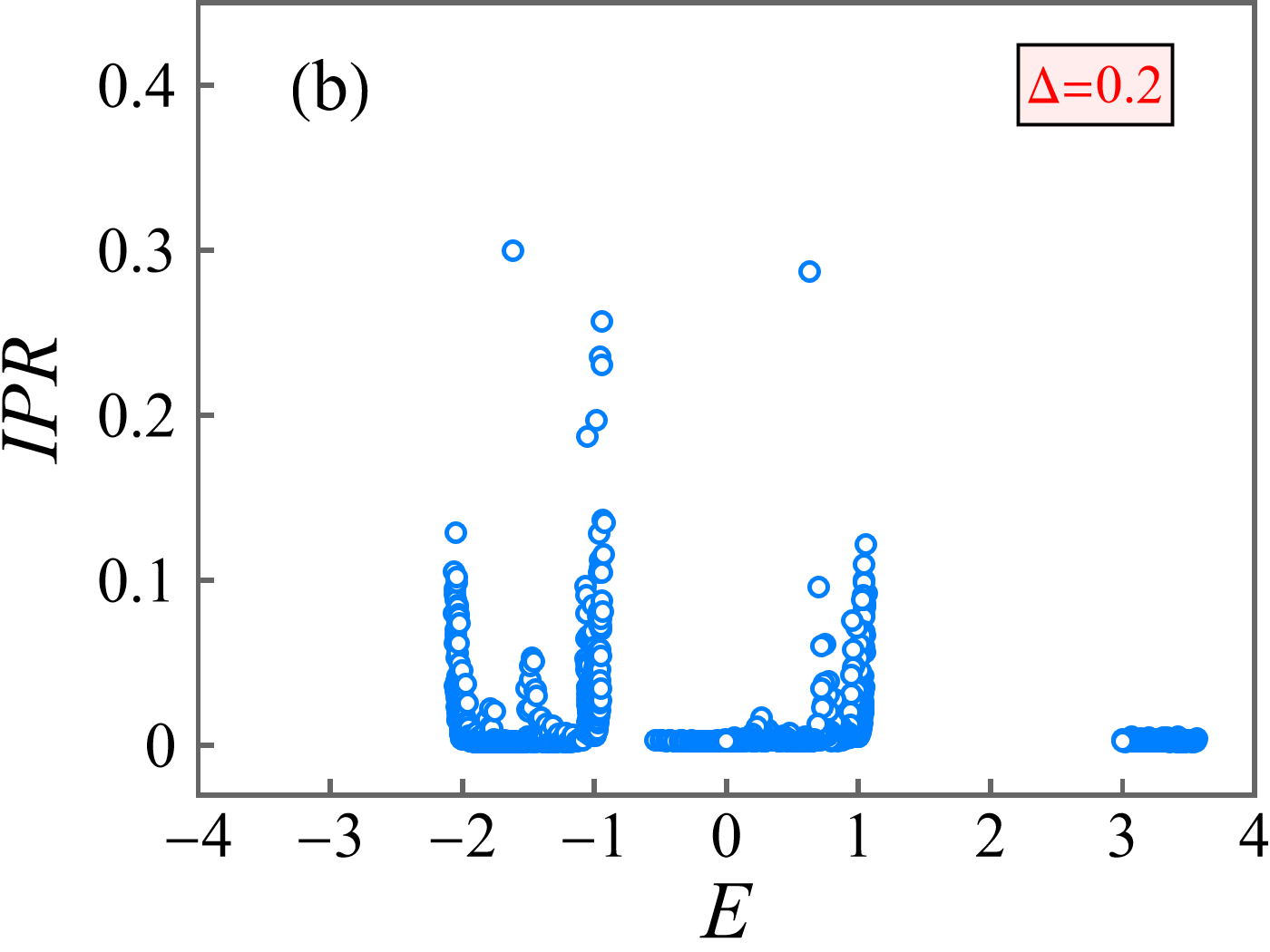} \\
\vskip 0.2cm
\includegraphics[clip, width=0.49\columnwidth]{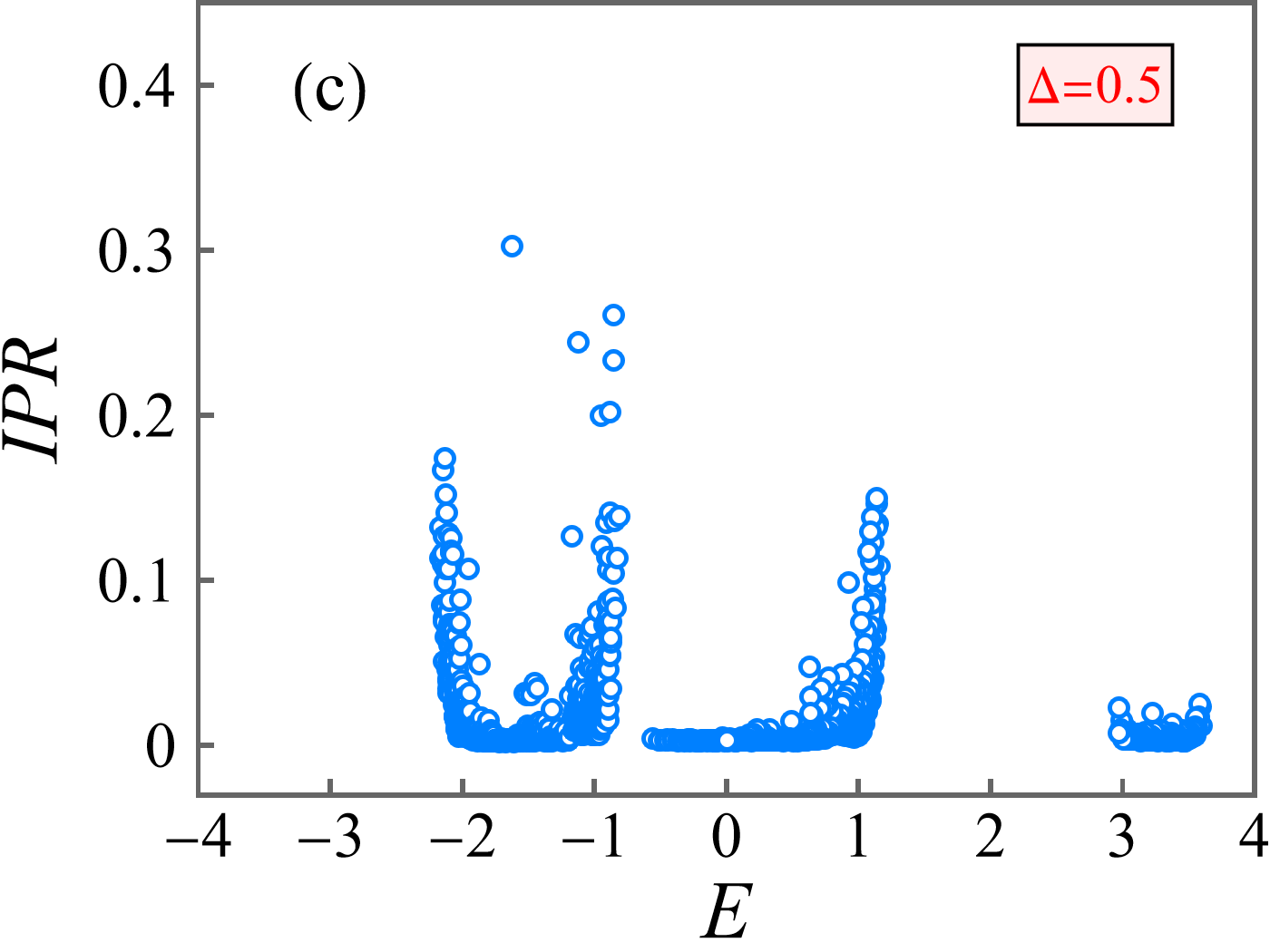}
\includegraphics[clip, width=0.49\columnwidth]{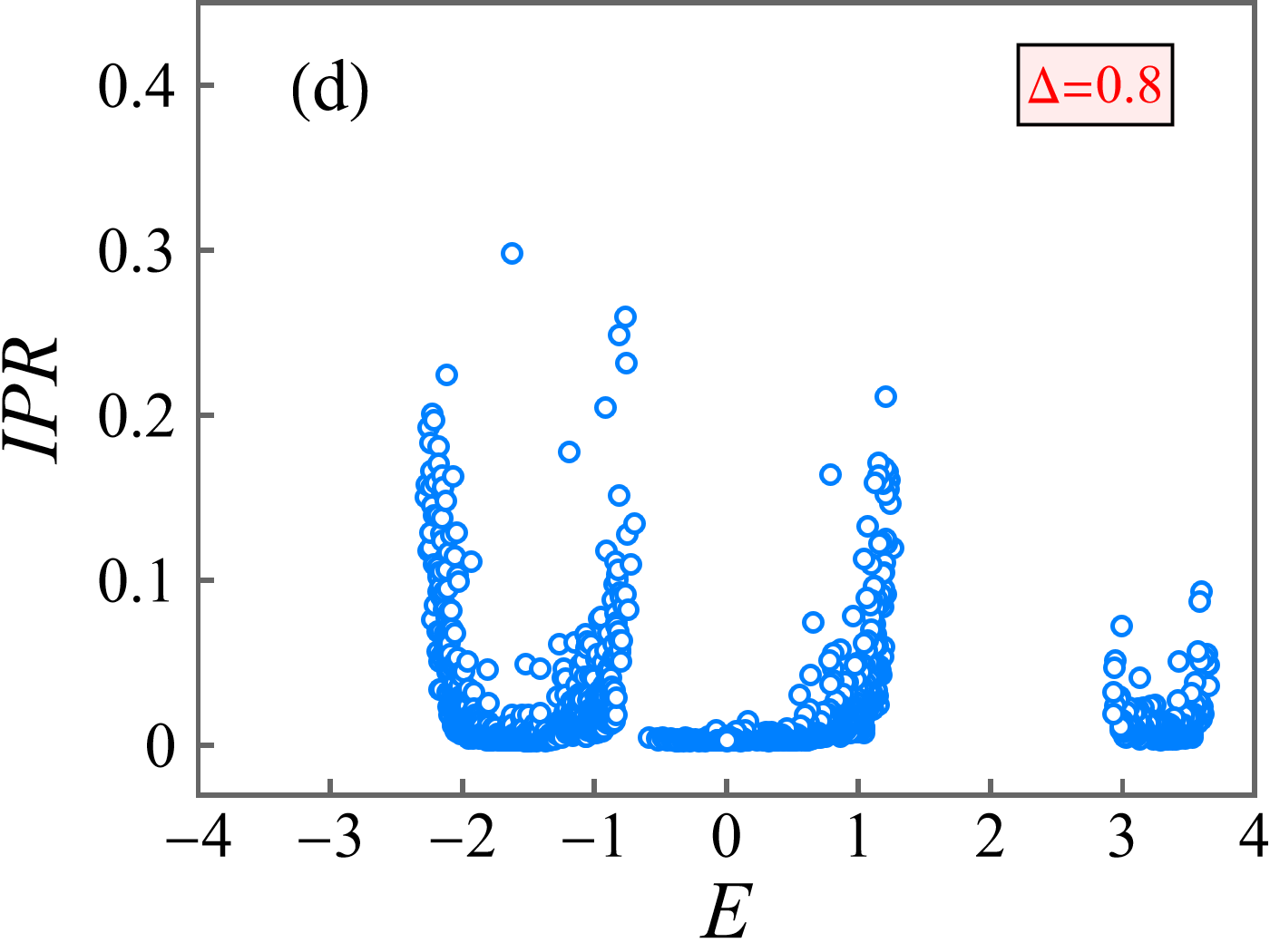}
\caption{The variation of the inverse participation ratio (\textit{IPR}) as a 
function of the energy ($E$) for a system-size with $15 \times 15$ unit 
cells for various random onsite disorder strengths, \textit{viz.}, (a) $\Delta=0$, 
(b) $\Delta=0.2$, (c) $\Delta=0.5$, and (d) $\Delta=0.8$ (measured in units 
of $t$). The value of $\Phi=0$.}
\label{fig:IPR}
\end{figure}
In this appendix, we have calculated the inverse participation ratio (\textit{IPR}) in order to 
quantify the localization-delocalization character of the single-particle states for the diamond-dodecagon lattice model. 
It is defined as follows: 
\begin{equation}
\mathit{IPR} = \dfrac{\sum\limits_{i=1}^{\mathcal{N}} |\psi_i|^4}
					{\Big(\sum\limits_{i=1}^{\mathcal{N}}|\psi_i|^2\Big)^2}, 
\label{eq:IPR}
\end{equation}
where the index $i$ runs over the total number of lattice sites $\mathcal{N}$ for given system-size. We have shown 
the variation of the \textit{IPR} for various onsite disorder strengths in Fig.~\ref{fig:IPR}. We emphasize on the 
fact that, smaller values of \textit{IPR} represent the extended states, while the large values of \textit{IPR} 
indicate the localized states. From the ADOS plots in Fig.~\ref{fig:DOS}, it was apparently found that the 
FB-induced peaks were smeared down as we increase the value of the onsite disorder strength $\Delta$; However, 
from the plots of the \textit{IPR} in Fig.~\ref{fig:IPR}, we observe that the localized character of the FB states 
still persist even if one increases the disorder strength $\Delta$. This is an indication towards the robustness of 
the FB states appearing in this lattice model.    

\end{document}